\documentclass[aps,prd,twocolumn,superscriptaddress,letterpaper,footinbib,nopreprintnumbers]{revtex4-1}

\usepackage[usenames,dvipsnames]{color}
\usepackage{graphicx}
\usepackage{float}
\usepackage{amsmath}
\usepackage{multirow}
\usepackage{ifthen}
\usepackage{blindtext}
\usepackage{svn-multi}
\usepackage[normalem]{ulem}
\usepackage{hyperref}
\usepackage{breakurl}
\usepackage{etoolbox}
\usepackage{fancyhdr}
\usepackage{import}
\newtoggle{arXiv}
\toggletrue{arXiv}  

\def\laq{\gtrsim}

\renewcommand{\today}{\number\day\space\ifcase\month\or
  January\or February\or March\or April\or May\or June\or
  July\or August\or September\or October\or November\or December\fi
  \space\number\year}

\newcommand{\makevisible}[1]{\textbf{\color{cyan}#1}}
\newcommand{\switch}[1]{%
  \ifthenelse{\equal{#1}{0}}{\renewcommand{\makevisible}[1]{}}{}}
\switch{0}

\newcommand{\macro}[1]{\textcolor{black}{#1}}

\newcommand{\Msun}{\ensuremath{{M}_\odot}}
\newcommand{\MMax}{\ensuremath{M_\mathrm{max}}}
\newcommand{\MMin}{\ensuremath{m_\mathrm{min}}}

\newcommand{\rateintrvcombimf}{\ensuremath{R = \macro{{103}^{+110}_{-63}}~\mathrm{Gpc^{-3}\,yr}^{-1}}}
\newcommand{\rateintrcomblog}{\ensuremath{R = \macro{{32}^{+33}_{-20}}~\mathrm{Gpc^{-3}\,yr}^{-1}}}
\newcommand{\rateintrbracket}{\ensuremath{\macro{12\text{--}213}~\mathrm{Gpc^{-3}\,yr}^{-1}}}
\newcommand{\rateintrbracketlow}{\ensuremath{\macro{12}~\mathrm{Gpc^{-3}\,yr}^{-1}}}

\newcommand{\pmexpintr}{\ensuremath{\macro{{2.3}^{+1.3}_{-1.4}}}}

\newcommand{\EventName}{\macro{GW170104}}
\newcommand{\EventTime}{\macro{10:11:58.6 UTC}}
\newcommand{\EventDate}{\macro{January 4, 2017}}
\newcommand{\OTwoChunkThreeLength}{\macro{5.5}}
\newcommand{\OTwoChunkThreeStart}{\macro{January 4, 2017}}
\newcommand{\OTwoChunkThreeEnd}{\macro{January 22, 2017}}

\newcommand{\EventNetworkSNR}{\macro{13}}
\newcommand{\EventFAR}{\macro{1 in 70,000 years}}

\newcommand{\OtwoJanBBHCalAmplUncert}{\macro{\ensuremath{5\%}}}
\newcommand{\OtwoJanBBHCalPhaseUncert}{\macro{\ensuremath{3^\circ}}}

\newcommand{\seobnrMTOTobsPerihelion}{\macro{\ensuremath{59.9_{-6.9}^{+5.7}}}} 
\newcommand{\imrppMTOTobsPerihelion}{\macro{\ensuremath{60.0_{-5.8}^{+5.7}}}} 
\newcommand{\MTOTobsPerihelion}{\macro{\ensuremath{59.9_{-6.5\pm1.0}^{+5.7\pm0.1}}}} 

\newcommand{\seobnrMTOTSPerihelion}{\macro{\ensuremath{50.4_{-5.1}^{+5.9}}}} 
\newcommand{\imrppMTOTSPerihelion}{\macro{\ensuremath{51.0_{-4.9}^{+5.8}}}} 
\newcommand{\MTOTSPerihelion}{\macro{\ensuremath{50.7_{-5.0\pm0.6}^{+5.9\pm0.5}}}} 

\newcommand{\seobnrMCobsPerihelion}{\macro{\ensuremath{25.2_{-4.2}^{+2.4}}}} 
\newcommand{\imrppMCobsPerihelion}{\macro{\ensuremath{24.9_{-3.5}^{+2.5}}}} 
\newcommand{\MCobsPerihelion}{\macro{\ensuremath{25.1_{-3.9\pm0.4}^{+2.5\pm0.2}}}} 

\newcommand{\seobnrMCSPerihelion}{\macro{\ensuremath{21.1_{-2.8}^{+2.3}}}} 
\newcommand{\imrppMCSPerihelion}{\macro{\ensuremath{21.1_{-2.5}^{+2.4}}}} 
\newcommand{\MCSPerihelion}{\macro{\ensuremath{21.1_{-2.7\pm0.3}^{+2.4\pm0.1}}}} 

\newcommand{\seobnrMONEobsPerihelion}{\macro{\ensuremath{36.3_{-6.5}^{+8.7}}}} 
\newcommand{\imrppMONEobsPerihelion}{\macro{\ensuremath{37.1_{-7.0}^{+9.2}}}} 
\newcommand{\MONEobsPerihelion}{\macro{\ensuremath{36.7_{-6.8\pm0.4}^{+9.0\pm1.2}}}} 

\newcommand{\seobnrMONESPerihelion}{\macro{\ensuremath{30.7_{-5.9}^{+8.1}}}} 
\newcommand{\imrppMONESPerihelion}{\macro{\ensuremath{31.6_{-6.3}^{+8.8}}}} 
\newcommand{\MONESPerihelion}{\macro{\ensuremath{31.2_{-6.0\pm0.4}^{+8.4\pm1.3}}}} 

\newcommand{\seobnrMTWOobsPerihelion}{\macro{\ensuremath{23.3_{-7.8}^{+5.9}}}} 
\newcommand{\imrppMTWOobsPerihelion}{\macro{\ensuremath{22.6_{-7.2}^{+6.5}}}} 
\newcommand{\MTWOobsPerihelion}{\macro{\ensuremath{22.9_{-7.5\pm0.1}^{+6.2\pm0.2}}}} 

\newcommand{\seobnrMTWOSPerihelion}{\macro{\ensuremath{19.6_{-6.2}^{+5.0}}}} 
\newcommand{\imrppMTWOSPerihelion}{\macro{\ensuremath{19.2_{-5.7}^{+5.4}}}} 
\newcommand{\MTWOSPerihelion}{\macro{\ensuremath{19.4_{-5.9\pm0.0}^{+5.3\pm0.1}}}} 

\newcommand{\seobnrMASSRATIOPerihelion}{\macro{\ensuremath{0.64_{-0.27}^{+0.31}}}} 
\newcommand{\imrppMASSRATIOPerihelion}{\macro{\ensuremath{0.60_{-0.26}^{+0.33}}}} 
\newcommand{\MASSRATIOPerihelion}{\macro{\ensuremath{0.62_{-0.26\pm0.02}^{+0.32\pm0.01}}}} 

\newcommand{\seobnrCHIEFFPerihelion}{\macro{\ensuremath{-0.11_{-0.33}^{+0.21}}}} 
\newcommand{\imrppCHIEFFPerihelion}{\macro{\ensuremath{-0.12_{-0.26}^{+0.20}}}} 
\newcommand{\CHIEFFPerihelion}{\macro{\ensuremath{-0.12_{-0.30\pm0.05}^{+0.21\pm0.01}}}} 

\newcommand{\seobnrCHIPPerihelion}{\macro{\ensuremath{0.40_{-0.30}^{+0.42}}}} 
\newcommand{\imrppCHIPPerihelion}{\macro{\ensuremath{0.42_{-0.29}^{+0.41}}}} 
\newcommand{\CHIPPerihelion}{\macro{\ensuremath{0.41_{-0.30\pm0.02}^{+0.41\pm0.00}}}} 

\newcommand{\seobnrSPINONEPerihelion}{\macro{\ensuremath{0.44_{-0.40}^{+0.48}}}} 
\newcommand{\imrppSPINONEPerihelion}{\macro{\ensuremath{0.45_{-0.40}^{+0.44}}}} 
\newcommand{\SPINONEPerihelion}{\macro{\ensuremath{0.45_{-0.40\pm0.01}^{+0.46\pm0.02}}}} 

\newcommand{\seobnrSPINTWOPerihelion}{\macro{\ensuremath{0.44_{-0.40}^{+0.48}}}} 
\newcommand{\imrppSPINTWOPerihelion}{\macro{\ensuremath{0.50_{-0.45}^{+0.44}}}} 
\newcommand{\SPINTWOPerihelion}{\macro{\ensuremath{0.47_{-0.43\pm0.01}^{+0.46\pm0.01}}}} 

\newcommand{\seobnrDISTANCEPerihelion}{\macro{\ensuremath{910_{-410}^{+470}}}} 
\newcommand{\imrppDISTANCEPerihelion}{\macro{\ensuremath{860_{-370}^{+410}}}} 
\newcommand{\DISTANCEPerihelion}{\macro{\ensuremath{880_{-390\pm10}^{+450\pm90}}}} 

\newcommand{\seobnrREDSHIFTPerihelion}{\macro{\ensuremath{0.182_{-0.078}^{+0.081}}}} 
\newcommand{\imrppREDSHIFTPerihelion}{\macro{\ensuremath{0.173_{-0.071}^{+0.072}}}} 
\newcommand{\REDSHIFTPerihelion}{\macro{\ensuremath{0.176_{-0.074\pm0.001}^{+0.078\pm0.015}}}} 

\newcommand{\imrppTIMEDELAYHLPerihelion}{\macro{\ensuremath{3.0_{-0.5}^{+0.4}}}} 

\newcommand{\seobnrPEOPTIMALSNRPerihelion}{\macro{\ensuremath{12.9_{-1.7}^{+1.7}}}} 
\newcommand{\imrppPEOPTIMALSNRPerihelion}{\macro{\ensuremath{13.0_{-1.7}^{+1.7}}}} 
\newcommand{\PEOPTIMALSNRPerihelion}{\macro{\ensuremath{13.0_{-1.7\pm0.0}^{+1.7\pm0.0}}}} 

\newcommand{\seobnrPEMATCHSNRPerihelion}{\macro{\ensuremath{13.3_{-0.3}^{+0.2}}}} 
\newcommand{\imrppPEMATCHSNRPerihelion}{\macro{\ensuremath{13.3_{-0.3}^{+0.2}}}} 
\newcommand{\PEMATCHSNRPerihelion}{\macro{\ensuremath{13.3_{-0.3\pm0.0}^{+0.2\pm0.0}}}} 

\newcommand{\MTOTSCOMPACTPerihelion}{\macro{\ensuremath{50.7_{-5.0}^{+5.9}}}} 
\newcommand{\MCSCOMPACTPerihelion}{\macro{\ensuremath{21.1_{-2.7}^{+2.4}}}} 
\newcommand{\MONESCOMPACTPerihelion}{\macro{\ensuremath{31.2_{-6.0}^{+8.4}}}} 
\newcommand{\MTWOSCOMPACTPerihelion}{\macro{\ensuremath{19.4_{-5.9}^{+5.3}}}} 
\newcommand{\CHIEFFCOMPACTPerihelion}{\macro{\ensuremath{-0.12_{-0.30}^{+0.21}}}} 
\newcommand{\DISTANCECOMPACTPerihelion}{\macro{\ensuremath{880_{-390}^{+450}}}} 
\newcommand{\REDSHIFTCOMPACTPerihelion}{\macro{\ensuremath{0.18_{-0.07}^{+0.08}}}} 

\newcommand{\MTOTSSIMPLEPerihelion}{\macro{\ensuremath{50}}} 
\newcommand{\DISTANCELTYRSIMPLEPerihelion}{\macro{\ensuremath{3}}} 
\newcommand{\imrppTIMEDELAYHLSIMPLEPerihelion}{\macro{\ensuremath{3}}} 

\newcommand{\seobnrSPINFINALinplanePerihelion}{\macro{\ensuremath{0.64_{-0.20}^{+0.09}}}} 
\newcommand{\imrppSPINFINALinplanePerihelion}{\macro{\ensuremath{0.63_{-0.19}^{+0.10}}}} 
\newcommand{\SPINFINALinplanePerihelion}{\macro{\ensuremath{0.64_{-0.20\pm0.00}^{+0.09\pm0.01}}}} 

\newcommand{\seobnrMFINALobsavgPerihelion}{\macro{\ensuremath{57.4_{-6.1}^{+5.4}}}} 
\newcommand{\imrppMFINALobsavgPerihelion}{\macro{\ensuremath{57.6_{-5.3}^{+5.6}}}} 
\newcommand{\MFINALobsavgPerihelion}{\macro{\ensuremath{57.5_{-5.8\pm0.8}^{+5.5\pm0.3}}}} 

\newcommand{\seobnrMFINALSavgPerihelion}{\macro{\ensuremath{48.4_{-4.7}^{+5.7}}}} 
\newcommand{\imrppMFINALSavgPerihelion}{\macro{\ensuremath{49.0_{-4.6}^{+5.7}}}} 
\newcommand{\MFINALSavgPerihelion}{\macro{\ensuremath{48.7_{-4.6\pm0.6}^{+5.7\pm0.6}}}} 

\newcommand{\SPINFINALinplaneCOMPACTPerihelion}{\macro{\ensuremath{0.64_{-0.20}^{+0.09}}}} 
\newcommand{\MFINALSavgCOMPACTPerihelion}{\macro{\ensuremath{48.7_{-4.6}^{+5.7}}}} 

\newcommand{\THETAJNPROBPerihelion}{\macro{\ensuremath{0.62}}} 

\newcommand{\seobnrSPINONELIMITPerihelion}{\macro{\ensuremath{0.86}}} 
\newcommand{\imrppSPINONELIMITPerihelion}{\macro{\ensuremath{0.82}}} 
\newcommand{\SPINONELIMITSYSPerihelion}{\macro{\ensuremath{0.84\pm0.03}}} 
\newcommand{\seobnrSPINTWOLIMITPerihelion}{\macro{\ensuremath{0.86}}} 
\newcommand{\imrppSPINTWOLIMITPerihelion}{\macro{\ensuremath{0.89}}} 
\newcommand{\SPINTWOLIMITSYSPerihelion}{\macro{\ensuremath{0.88\pm0.02}}} 

\newcommand{\seobnrCHIEFFLIMITPerihelion}{\macro{\ensuremath{0.05}}} 
\newcommand{\imrppCHIEFFLIMITPerihelion}{\macro{\ensuremath{0.04}}} 
\newcommand{\CHIEFFLIMITSYSPerihelion}{\macro{\ensuremath{0.04\pm0.01}}} 

\newcommand{\seobnrCHIPLIMITPerihelion}{\macro{\ensuremath{0.74}}} 
\newcommand{\imrppCHIPLIMITPerihelion}{\macro{\ensuremath{0.74}}} 
\newcommand{\CHIPLIMITSYSPerihelion}{\macro{\ensuremath{0.74\pm0.01}}} 

\newcommand{\CHIEFFPROBPerihelion}{\macro{\ensuremath{0.18}}} 
\newcommand{\CHIEFFNEGPROBPerihelion}{\macro{\ensuremath{0.82}}} 

\newcommand{\ALIGNEDPROBPerihelion}{\macro{\ensuremath{0.05}}} 

\newcommand{\seobnrMASSRATIOLIMITPerihelion}{\macro{\ensuremath{0.42}}} 
\newcommand{\imrppMASSRATIOLIMITPerihelion}{\macro{\ensuremath{0.40}}} 
\newcommand{\MASSRATIOLIMITSYSPerihelion}{\macro{\ensuremath{0.41\pm0.02}}} 

\newcommand{\TILTANGLEONEPROBPerihelion}{\macro{\ensuremath{0.04}}} 
\newcommand{\TILTANGLETWOPROBPerihelion}{\macro{\ensuremath{0.08}}} 

\newcommand{\seobnrERADPerihelion}{\macro{\ensuremath{2.1_{-0.8}^{+0.5}}}} 
\newcommand{\imrppERADPerihelion}{\macro{\ensuremath{2.0_{-0.7}^{+0.6}}}} 
\newcommand{\ERADPerihelion}{\macro{\ensuremath{2.0_{-0.7\pm0.0}^{+0.6\pm0.0}}}} 

\newcommand{\ERADCOMPACTPerihelion}{\macro{\ensuremath{2.0_{-0.7}^{+0.6}}}} 
\newcommand{\LPEAKCOMPACTPerihelion}{\macro{\ensuremath{3.1_{-1.3}^{+0.7} \times 10^{56}}}} 

\newcommand{\PESKYNINTYPerihelion}{\macro{\ensuremath{1200~\mathrm{deg^2}}}} 

\newcommand{\SKYCALIBNINTYimrppPerihelion}{\macro{\ensuremath{2\%}}} 

\newcommand{\CHIEFFKLPerihelion}{\macro{\ensuremath{0.36}}} 
\newcommand{\CHIPKLPerihelion}{\macro{\ensuremath{0.03}}} 

\newcommand{\burstOverlap}{\macro{\ensuremath{87}}} 
\newcommand{\burstEventFAR}{\macro{$\sim$ 1 in 20,000 years}} 

\usepackage{amssymb}

\newcommand{\IMRTRANSITIONFREQPerihelion}{\macro{\ensuremath{143}}} 

\newcommand{\LAMBDAMGPerihelion}{\macro{\ensuremath{1.5\times 10^{13}}}} 
\newcommand{\LAMBDAMGGWFIRSTEVENT}{\macro{\ensuremath{1.0\times 10^{13}}}} 
\newcommand{\MASSMGCumulative}{\macro{\ensuremath{7.7\times 10^{-23}}}} 
\newcommand{\LAMBDAMGCumulative}{\macro{\ensuremath{1.6\times 10^{13}}}} 

\newcommand{\lora}{\ensuremath{A}} 

\newcommand{\CHIEFFPROB}{\macro{\ensuremath{0.23}}} 

\newcommand{\CHIEFFPROBSecondMonday}{\macro{\ensuremath{0.59}}} 

\newcommand{\SPINMAXMINBoxing}{\macro{\ensuremath{0.2}}} 

\newcommand{\CHIEFFMAGLIMITBoxing}{\macro{\ensuremath{0.35}}} 
\newtoggle{fullauthorlist}
\toggletrue{fullauthorlist}

\newtoggle{endauthorlist}
\toggletrue{endauthorlist}

\begin{document}

\title{GW170104: Observation of a 50-Solar-Mass Binary Black Hole Coalescence at Redshift~0.2}


\iftoggle{endauthorlist}{
  %
  %
  \let\mymaketitle\maketitle
  \let\myauthor\author
  \let\myaffiliation\affiliation
  \author{The LIGO Scientific Collaboration and the Virgo Collaboration}
}{
  %
  %
  \iftoggle{fullauthorlist}{



\author{%
B.~P.~Abbott,$^{1}$  
R.~Abbott,$^{1}$  
T.~D.~Abbott,$^{2}$  
F.~Acernese,$^{3,4}$ 
K.~Ackley,$^{5}$  
C.~Adams,$^{6}$  
T.~Adams,$^{7}$ 
P.~Addesso,$^{8}$  
R.~X.~Adhikari,$^{1}$  
V.~B.~Adya,$^{9}$
% ...
\\
\medskip
(LIGO Scientific Collaboration and Virgo Collaboration) 
\\
\medskip
{{}$^{*}$Deceased, March 2016. }%
{{}$^{\sharp}$Deceased, March 2017. }%
{${}^{\dag}$Deceased, February 2017. }%
{${}^{\ddag}$Deceased, December 2016. }%
}\noaffiliation
\affiliation {LIGO, California Institute of Technology, Pasadena, CA 91125, USA }
\affiliation {Louisiana State University, Baton Rouge, LA 70803, USA }
\affiliation {Universit\`a di Salerno, Fisciano, I-84084 Salerno, Italy }
\affiliation {INFN, Sezione di Napoli, Complesso Universitario di Monte S.Angelo, I-80126 Napoli, Italy }
\affiliation {University of Florida, Gainesville, FL 32611, USA }
\affiliation {LIGO Livingston Observatory, Livingston, LA 70754, USA }
\affiliation {Laboratoire d'Annecy-le-Vieux de Physique des Particules (LAPP), Universit\'e Savoie Mont Blanc, CNRS/IN2P3, F-74941 Annecy, France }
\affiliation {University of Sannio at Benevento, I-82100 Benevento, Italy and INFN, Sezione di Napoli, I-80100 Napoli, Italy }
\affiliation {Albert-Einstein-Institut, Max-Planck-Institut f\"ur Gravi\-ta\-tions\-physik, D-30167 Hannover, Germany }
\affiliation {The University of Mississippi, University, MS 38677, USA }
\affiliation {NCSA, University of Illinois at Urbana-Champaign, Urbana, IL 61801, USA }
\affiliation {University of Cambridge, Cambridge CB2 1TN, United Kingdom }
\affiliation {Nikhef, Science Park, 1098 XG Amsterdam, Netherlands }
\affiliation {LIGO, Massachusetts Institute of Technology, Cambridge, MA 02139, USA }
\affiliation {Instituto Nacional de Pesquisas Espaciais, 12227-010 S\~{a}o Jos\'{e} dos Campos, S\~{a}o Paulo, Brazil }
\affiliation {Gran Sasso Science Institute (GSSI), I-67100 L'Aquila, Italy }
\affiliation {INFN, Sezione di Roma Tor Vergata, I-00133 Roma, Italy }
\affiliation {Inter-University Centre for Astronomy and Astrophysics, Pune 411007, India }
\affiliation {International Centre for Theoretical Sciences, Tata Institute of Fundamental Research, Bengaluru 560089, India }
\affiliation {University of Wisconsin-Milwaukee, Milwaukee, WI 53201, USA }
\affiliation {Leibniz Universit\"at Hannover, D-30167 Hannover, Germany }
\affiliation {Universit\`a di Pisa, I-56127 Pisa, Italy }
\affiliation {INFN, Sezione di Pisa, I-56127 Pisa, Italy }
\affiliation {OzGrav, Australian National University, Canberra, Australian Capital Territory 0200, Australia }
\affiliation {Laboratoire des Mat\'eriaux Avanc\'es (LMA), CNRS/IN2P3, F-69622 Villeurbanne, France }
\affiliation {LAL, Univ. Paris-Sud, CNRS/IN2P3, Universit\'e Paris-Saclay, F-91898 Orsay, France }
\affiliation {California State University Fullerton, Fullerton, CA 92831, USA }
\affiliation {European Gravitational Observatory (EGO), I-56021 Cascina, Pisa, Italy }
\affiliation {Chennai Mathematical Institute, Chennai 603103, India }
\affiliation {Universit\`a di Roma Tor Vergata, I-00133 Roma, Italy }
\affiliation {Universit\"at Hamburg, D-22761 Hamburg, Germany }
\affiliation {INFN, Sezione di Roma, I-00185 Roma, Italy }
\affiliation {Embry-Riddle Aeronautical University, Prescott, AZ 86301, USA }
\affiliation {Albert-Einstein-Institut, Max-Planck-Institut f\"ur Gravitations\-physik, D-14476 Potsdam-Golm, Germany }
\affiliation {APC, AstroParticule et Cosmologie, Universit\'e Paris Diderot, CNRS/IN2P3, CEA/Irfu, Observatoire de Paris, Sorbonne Paris Cit\'e, F-75205 Paris Cedex 13, France }
\affiliation {Korea Institute of Science and Technology Information, Daejeon 34141, Korea }
\affiliation {West Virginia University, Morgantown, WV 26506, USA }
\affiliation {Center for Gravitational Waves and Cosmology, West Virginia University, Morgantown, WV 26505, USA }
\affiliation {Universit\`a di Perugia, I-06123 Perugia, Italy }
\affiliation {INFN, Sezione di Perugia, I-06123 Perugia, Italy }
\affiliation {Syracuse University, Syracuse, NY 13244, USA }
\affiliation {University of Minnesota, Minneapolis, MN 55455, USA }
\affiliation {SUPA, University of Glasgow, Glasgow G12 8QQ, United Kingdom }
\affiliation {LIGO Hanford Observatory, Richland, WA 99352, USA }
\affiliation {Wigner RCP, RMKI, H-1121 Budapest, Konkoly Thege Mikl\'os \'ut 29-33, Hungary }
\affiliation {Columbia University, New York, NY 10027, USA }
\affiliation {Stanford University, Stanford, CA 94305, USA }
\affiliation {Universit\`a di Camerino, Dipartimento di Fisica, I-62032 Camerino, Italy }
\affiliation {Universit\`a di Padova, Dipartimento di Fisica e Astronomia, I-35131 Padova, Italy }
\affiliation {INFN, Sezione di Padova, I-35131 Padova, Italy }
\affiliation {MTA E\"otv\"os University, ``Lendulet'' Astrophysics Research Group, Budapest 1117, Hungary }
\affiliation {Nicolaus Copernicus Astronomical Center, Polish Academy of Sciences, 00-716, Warsaw, Poland }
\affiliation {University of Birmingham, Birmingham B15 2TT, United Kingdom }
\affiliation {Universit\`a degli Studi di Genova, I-16146 Genova, Italy }
\affiliation {INFN, Sezione di Genova, I-16146 Genova, Italy }
\affiliation {RRCAT, Indore MP 452013, India }
\affiliation {Faculty of Physics, Lomonosov Moscow State University, Moscow 119991, Russia }
\affiliation {SUPA, University of the West of Scotland, Paisley PA1 2BE, United Kingdom }
\affiliation {Caltech CaRT, Pasadena, CA 91125, USA }
\affiliation {OzGrav, University of Western Australia, Crawley, Western Australia 6009, Australia }
\affiliation {Department of Astrophysics/IMAPP, Radboud University Nijmegen, P.O. Box 9010, 6500 GL Nijmegen, Netherlands }
\affiliation {Artemis, Universit\'e C\^ote d'Azur, Observatoire C\^ote d'Azur, CNRS, CS 34229, F-06304 Nice Cedex 4, France }
\affiliation {Institut de Physique de Rennes, CNRS, Universit\'e de Rennes 1, F-35042 Rennes, France }
\affiliation {Washington State University, Pullman, WA 99164, USA }
\affiliation {Universit\`a degli Studi di Urbino ``Carlo Bo'', I-61029 Urbino, Italy }
\affiliation {INFN, Sezione di Firenze, I-50019 Sesto Fiorentino, Firenze, Italy }
\affiliation {University of Oregon, Eugene, OR 97403, USA }
\affiliation {Laboratoire Kastler Brossel, UPMC-Sorbonne Universit\'es, CNRS, ENS-PSL Research University, Coll\`ege de France, F-75005 Paris, France }
\affiliation {Carleton College, Northfield, MN 55057, USA }
\affiliation {Astronomical Observatory Warsaw University, 00-478 Warsaw, Poland }
\affiliation {VU University Amsterdam, 1081 HV Amsterdam, Netherlands }
\affiliation {University of Maryland, College Park, MD 20742, USA }
\affiliation {Center for Relativistic Astrophysics and School of Physics, Georgia Institute of Technology, Atlanta, GA 30332, USA }
\affiliation {Universit\'e Claude Bernard Lyon 1, F-69622 Villeurbanne, France }
\affiliation {Universit\`a di Napoli ``Federico II'', Complesso Universitario di Monte S.~Angelo, I-80126 Napoli, Italy }
\affiliation {NASA Goddard Space Flight Center, Greenbelt, MD 20771, USA }
\affiliation {RESCEU, University of Tokyo, Tokyo, 113-0033, Japan }
\affiliation {OzGrav, University of Adelaide, Adelaide, South Australia 5005, Australia }
\affiliation {Tsinghua University, Beijing 100084, China }
\affiliation {Texas Tech University, Lubbock, TX 79409, USA }
\affiliation {Kenyon College, Gambier, OH 43022, USA }
\affiliation {The Pennsylvania State University, University Park, PA 16802, USA }
\affiliation {National Tsing Hua University, Hsinchu City, 30013 Taiwan, Republic of China }
\affiliation {Charles Sturt University, Wagga Wagga, New South Wales 2678, Australia }
\affiliation {Canadian Institute for Theoretical Astrophysics, University of Toronto, Toronto, Ontario M5S 3H8, Canada }
\affiliation {University of Chicago, Chicago, IL 60637, USA }
\affiliation {Pusan National University, Busan 46241, Korea }
\affiliation {The Chinese University of Hong Kong, Shatin, NT, Hong Kong }
\affiliation {INAF, Osservatorio Astronomico di Padova, Vicolo dell'Osservatorio 5, I-35122 Padova, Italy }
\affiliation {INFN, Trento Institute for Fundamental Physics and Applications, I-38123 Povo, Trento, Italy }
\affiliation {Universit\`a di Roma ``La Sapienza'', I-00185 Roma, Italy }
\affiliation {Universit\'e Libre de Bruxelles, Brussels 1050, Belgium }
\affiliation {Sonoma State University, Rohnert Park, CA 94928, USA }
\affiliation {Montana State University, Bozeman, MT 59717, USA }
\affiliation {Center for Interdisciplinary Exploration \& Research in Astrophysics (CIERA), Northwestern University, Evanston, IL 60208, USA }
\affiliation {Cardiff University, Cardiff CF24 3AA, United Kingdom }
\affiliation {Universitat de les Illes Balears, IAC3---IEEC, E-07122 Palma de Mallorca, Spain }
\affiliation {The University of Texas Rio Grande Valley, Brownsville, TX 78520, USA }
\affiliation {Bellevue College, Bellevue, WA 98007, USA }
\affiliation {Institute for Plasma Research, Bhat, Gandhinagar 382428, India }
\affiliation {The University of Sheffield, Sheffield S10 2TN, United Kingdom }
\affiliation {University of Southern California Information Sciences Institute, Marina Del Rey, CA 90292, USA }
\affiliation {California State University, Los Angeles, 5151 State University Dr, Los Angeles, CA 90032, USA }
\affiliation {Universit\`a di Trento, Dipartimento di Fisica, I-38123 Povo, Trento, Italy }
\affiliation {Montclair State University, Montclair, NJ 07043, USA }
\affiliation {National Astronomical Observatory of Japan, 2-21-1 Osawa, Mitaka, Tokyo 181-8588, Japan }
\affiliation {Whitman College, 345 Boyer Avenue, Walla Walla, WA 99362 USA }
\affiliation {School of Mathematics, University of Edinburgh, Edinburgh EH9 3FD, United Kingdom }
\affiliation {University and Institute of Advanced Research, Gandhinagar Gujarat 382007, India }
\affiliation {IISER-TVM, CET Campus, Trivandrum Kerala 695016, India }
\affiliation {University of Szeged, D\'om t\'er 9, Szeged 6720, Hungary }
\affiliation {Tata Institute of Fundamental Research, Mumbai 400005, India }
\affiliation {INAF, Osservatorio Astronomico di Capodimonte, I-80131, Napoli, Italy }
\affiliation {University of Michigan, Ann Arbor, MI 48109, USA }
\affiliation {American University, Washington, D.C. 20016, USA }
\affiliation {Rochester Institute of Technology, Rochester, NY 14623, USA }
\affiliation {University of Bia{\l }ystok, 15-424 Bia{\l }ystok, Poland }
\affiliation {SUPA, University of Strathclyde, Glasgow G1 1XQ, United Kingdom }
\affiliation {University of Southampton, Southampton SO17 1BJ, United Kingdom }
\affiliation {University of Washington Bothell, 18115 Campus Way NE, Bothell, WA 98011, USA }
\affiliation {Institute of Applied Physics, Nizhny Novgorod, 603950, Russia }
\affiliation {Seoul National University, Seoul 08826, Korea }
\affiliation {Inje University Gimhae, South Gyeongsang 50834, Korea }
\affiliation {National Institute for Mathematical Sciences, Daejeon 34047, Korea }
\affiliation {NCBJ, 05-400 \'Swierk-Otwock, Poland }
\affiliation {Institute of Mathematics, Polish Academy of Sciences, 00656 Warsaw, Poland }
\affiliation {OzGrav, School of Physics \& Astronomy, Monash University, Clayton 3800, Victoria, Australia }
\affiliation {Hanyang University, Seoul 04763, Korea }
\affiliation {NASA Marshall Space Flight Center, Huntsville, AL 35811, USA }
\affiliation {ESPCI, CNRS, F-75005 Paris, France }
\affiliation {Southern University and A\&M College, Baton Rouge, LA 70813, USA }
\affiliation {OzGrav, University of Melbourne, Parkville, Victoria 3010, Australia }
\affiliation {College of William and Mary, Williamsburg, VA 23187, USA }
\affiliation {Indian Institute of Technology Madras, Chennai 600036, India }
\affiliation {IISER-Kolkata, Mohanpur, West Bengal 741252, India }
\affiliation {Scuola Normale Superiore, Piazza dei Cavalieri 7, I-56126 Pisa, Italy }
\affiliation {Universit\'e de Lyon, F-69361 Lyon, France }
\affiliation {Hobart and William Smith Colleges, Geneva, NY 14456, USA }
\affiliation {Janusz Gil Institute of Astronomy, University of Zielona G\'ora, 65-265 Zielona G\'ora, Poland }
\affiliation {University of Washington, Seattle, WA 98195, USA }
\affiliation {King's College London, University of London, London WC2R 2LS, United Kingdom }
\affiliation {Indian Institute of Technology, Gandhinagar Ahmedabad Gujarat 382424, India }
\affiliation {International Institute of Physics, Universidade Federal do Rio Grande do Norte, Natal RN 59078-970, Brazil }
\affiliation {Andrews University, Berrien Springs, MI 49104, USA }
\affiliation {Universit\`a di Siena, I-53100 Siena, Italy }
\affiliation {Trinity University, San Antonio, TX 78212, USA }
\affiliation {Abilene Christian University, Abilene, TX 79699, USA }


  }{
    \author{The LIGO Scientific Collaboration and the Virgo Collaboration}
  }
}

\begin{abstract}
\iftoggle{endauthorlist}{
}{
 \iftoggle{fullauthorlist}{
  \newpage
 }
}
We describe the observation of \EventName{}, a gravitational-wave signal produced by the coalescence of a pair of stellar-mass black holes.
The signal was measured on \EventDate{} at \EventTime{} by the twin advanced detectors of the Laser Interferometer Gravitational-Wave Observatory during their second observing run, with a network signal-to-noise ratio of 13 and a false alarm rate less than \EventFAR{}.
The inferred component black hole masses are $\MONESCOMPACTPerihelion{}\,\Msun$ and $\MTWOSCOMPACTPerihelion{}\,\Msun$ 
(at the $90\%$ credible level). 
The black hole spins are best constrained through measurement of the effective inspiral spin parameter, a mass-weighted combination of the spin components perpendicular to the orbital plane, 
$\chi_\mathrm{eff} = \CHIEFFCOMPACTPerihelion$. 
This result implies that spin configurations with both component spins positively aligned with the orbital angular momentum are disfavored. 
The source luminosity distance is $\DISTANCECOMPACTPerihelion~\mathrm{Mpc}$ corresponding to a redshift of  $z = \REDSHIFTCOMPACTPerihelion$. We constrain the magnitude of modifications to the gravitational-wave dispersion relation and perform null tests of general relativity.  Assuming that gravitons are dispersed in vacuum like massive particles, we bound the graviton mass to $m_g\leq\MASSMGCumulative~\mathrm{eV}/c^2$.  
In all cases, we find that \EventName{} is consistent with general relativity.
\end{abstract}

\maketitle

\section{Introduction}
\label{s:intro}

The first observing run of the Advanced Laser Interferometer Gravitational-Wave Observatory 
(LIGO)~\cite{TheLIGOScientific:2014jea} identified two binary black hole 
coalescence signals with high statistical significance, GW150914~\cite{GW150914-DETECTION} and 
GW151226~\cite{GW151226-DETECTION}, as well as a less significant candidate 
LVT151012~\cite{GW150914-CBC,O1:BBH}. These discoveries ushered in a new era of observational
astronomy, allowing us to investigate the astrophysics of binary black holes and test general 
relativity (GR) in ways that were previously inaccessible~\cite{GW150914-ASTRO,GW150914-TESTOFGR}. 
We now know that there is a population of binary black holes with component 
masses $\gtrsim 25\,\Msun$~\cite{GW150914-ASTRO,O1:BBH}, and that 
merger rates are high enough for us to expect more detections~\cite{GW150914-RATES,O1:BBH}. 

\begin{figure}
\vskip-10pt
 \centering
\includegraphics[width=1\columnwidth]{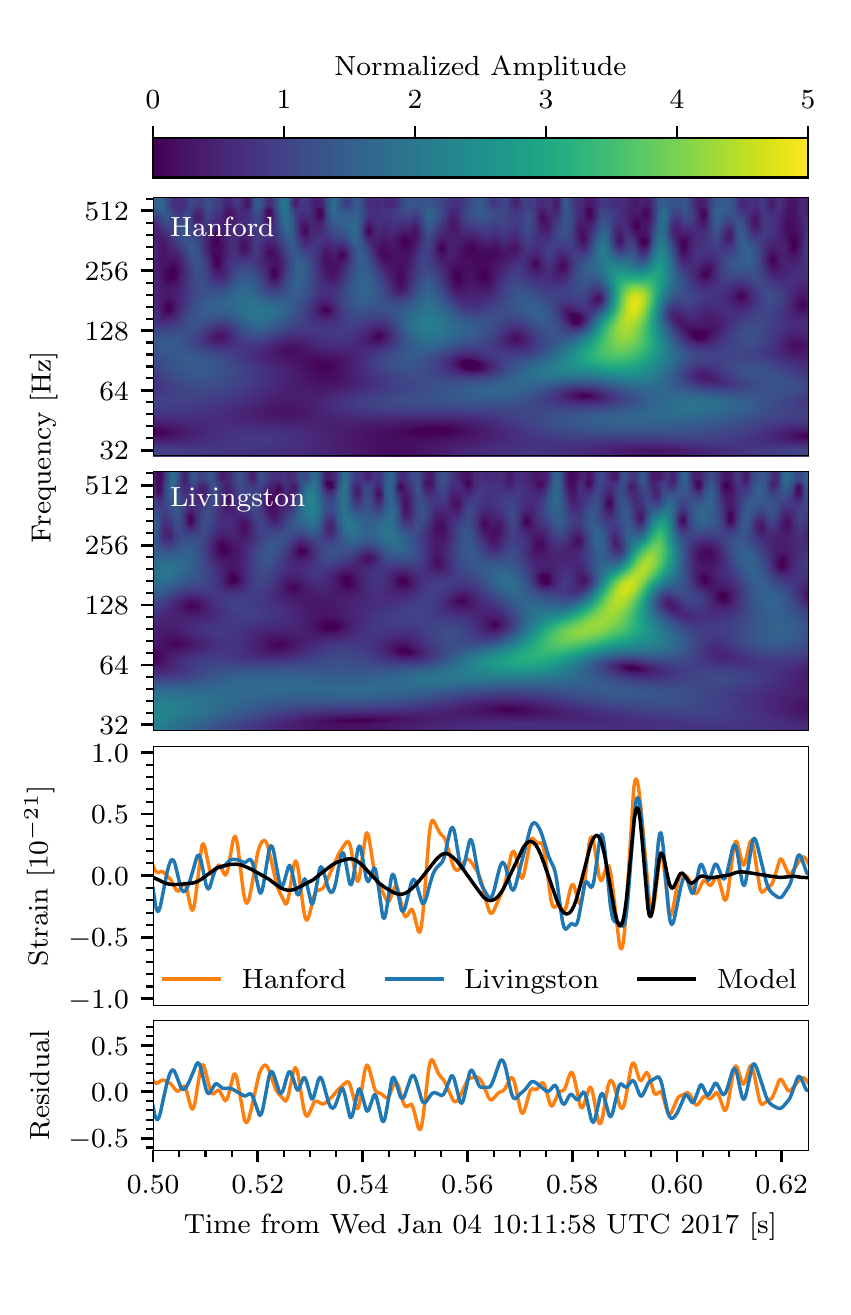}
\vskip-10pt
\caption{Time--frequency representation~\cite{Chatterji:2004qg} of strain data
from Hanford and Livingston detectors (top two panels) at the time of \EventName.  The data
begin at $1167559936.5$ GPS time. The third panel from top shows the time-series data
from each detector with a $30$--$350~\mathrm{Hz}$ bandpass filter, and band-reject filters 
to suppress strong instrumental spectral lines.  The
Livingston data have been shifted back by $3~\mathrm{ms}$ to account for the source's sky
location, and the sign of its amplitude has been inverted to account for the
detectors' different orientations.  The maximum-likelihood binary black hole
waveform given by the full-precession model (see Sec.~\ref{s:parameters}) is
shown in black.  The bottom panel shows the residuals between each data stream
and the maximum-likelihood waveform.  
} \label{fig:omega_scan}
\end{figure}
 
Advanced LIGO's second observing run began on November 30, 2016. 
On \EventDate{}, 
a gravitational-wave signal
was detected with high statistical significance. 
Figure~\ref{fig:omega_scan} shows a time--frequency representation of the 
data from the LIGO Hanford and Livingston detectors, with the 
signal \EventName{} visible as the characteristic chirp 
of a binary coalescence. 
Detailed analyses demonstrate 
that \EventName{} arrived at Hanford $\sim\imrppTIMEDELAYHLSIMPLEPerihelion~\mathrm{ms}$ before Livingston, 
and originated from the coalescence of two stellar-mass black holes at a luminosity distance of
$\sim\DISTANCELTYRSIMPLEPerihelion$ billion light-years. 
 
\EventName's source is a heavy binary black hole system, with a total mass of $\sim\MTOTSSIMPLEPerihelion\,\Msun$, 
suggesting formation in a sub-solar metallicity environment~\cite{GW150914-ASTRO}. 
Measurements of the black hole spins show a preference away from being (positively) aligned with the orbital angular momentum, 
but do not exclude zero spins. 
This is distinct from the case for GW151226, 
which had a strong preference 
for spins with positive projections along the orbital angular momentum~\cite{GW151226-DETECTION}. 
The inferred merger rate agrees with previous calculations~\cite{GW150914-RATES,O1:BBH}, 
and could potentially be explained by binary black holes 
forming through isolated binary evolution or dynamical interactions in dense stellar clusters~\cite{GW150914-ASTRO}. 

Gravitational-wave observations of binary black holes are the ideal means to test GR 
and its alternatives. 
They provide insight into regimes of strong-field gravity 
where velocities are relativistic and the spacetime is dynamic. 
The tests performed with the sources detected in the first observing run  
showed no evidence of departure from GR's predictions~\cite{GW150914-TESTOFGR,O1:BBH}; 
\EventName{} provides an opportunity to tighten these constraints. 
In addition to repeating tests performed in the first observing run, we also test for 
modifications to the gravitational-wave dispersion relation. 
Combining measurements from \EventName{} with our previous results, 
we obtain new gravitational-wave constraints on potential deviations from GR.

\section{Detectors and data quality}
\label{s:detectors}
The LIGO detectors measure gravitational-wave strain using two dual-recycled Fabry--Perot Michelson 
interferometers at the Hanford and Livingston observatories~\cite{GW150914-DETECTORS,TheLIGOScientific:2014jea}.  
After the first observing run, both LIGO detectors 
underwent commissioning
to reduce instrumental noise,
and to improve duty factor and data quality
\iftoggle{arXiv}{%
  (see Appendix~\ref{s:detectors_extra} in the \textit{Supplemental Material}~\cite{Supp-Mat}).
}{
  (see Sec.~I of the \textit{Supplemental Material}~\cite{Supp-Mat}).
}
For the Hanford detector, a high-power 
laser stage was introduced, and as the first step the laser power was
increased from $22~\mathrm{W}$ to $30~\mathrm{W}$ to reduce
shot noise~\cite{GW150914-DETECTORS} at high frequencies.
For the Livingston detector, the laser power 
was unchanged,
but there was a significant improvement
in low-frequency performance mainly due to the mitigation of scattered light noise. 

Calibration of the interferometers is performed by inducing test-mass motion 
using photon pressure from modulated calibration lasers~\cite{Karki:2016pht,GW150914-CALIBRATION}. 
The one-sigma calibration uncertainties for strain data
in both detectors for the times used in this analysis are better than 
\OtwoJanBBHCalAmplUncert{} in amplitude and \OtwoJanBBHCalPhaseUncert{} 
in phase over the frequency range $20$--$1024~\mathrm{Hz}$.

At the time of \EventName{}, both LIGO detectors were operating
with sensitivity 
typical of the observing run to date and were in an 
observation-ready state.  Investigations similar to the detection validation 
procedures for previous events~\cite{GW150914-DETECTION,GW150914-DETCHAR} found no evidence that instrumental or 
environmental disturbances contributed to \EventName{}.

\section{Searches}
\label{s:searches}
\EventName{} was first identified by inspection of 
low-latency triggers from Livingston data~\cite{Canton:2014ena, Usman:2015kfa, pycbc-github}.
An automated notification was not generated as the Hanford detector's calibration state was temporarily 
set incorrectly in the low-latency system. 
After it was manually determined that the calibration of both detectors was in a
nominal state, an alert with an initial source
localization~\cite{Singer:2015ema,Singer:2016eax} was 
distributed to collaborating astronomers~\cite{GCN20364} for the purpose of searching
for a transient counterpart. 
Twenty-eight groups of observers covered the parts of the sky localization using ground- and space-based instruments, 
spanning from $\gamma$ ray to radio frequencies as well as high-energy neutrinos~\cite{GCN-archive}.

Offline analyses are used to determine the significance of candidate events.
They benefit from improved calibration and refined data 
quality information that is unavailable to low-latency analyses~\cite{O1:BBH, GW150914-DETCHAR}.
The second observing run is divided into periods of two-detector
cumulative coincident observing time with $\laq 5$ days of data to measure the false alarm rate of
the search at the level where detections can be confidently claimed. 
Two independently designed matched filter analyses~\cite{Usman:2015kfa, Messick:2016aqy}
used \OTwoChunkThreeLength{} days of coincident data collected
from \OTwoChunkThreeStart{} to \OTwoChunkThreeEnd{}. 

These analyses search for binary coalescences over a range of possible masses
and by using discrete banks~\cite{Capano:2016dsf,
Brown:2012qf, Harry:2009ea, Ajith:2012mn, Sathyaprakash:1991mt, Owen:1998dk}
of waveform templates modeling binaries with component spins 
aligned or antialigned with the orbital angular momentum~\cite{Bohe:2016gbl}.
The searches can target binary black hole mergers with detector-frame total masses $2\,\Msun \le M^\mathrm{det} \lesssim 100$--$500\, \Msun$, 
and spin magnitudes up to $\sim0.99$.
The upper mass boundary of the bank is determined by imposing a
 lower limit on the duration of the template in the
 detectors' sensitive frequency band~\cite{DalCanton:2017}.
Candidate events must be found in both detectors by the same template
within $15~\mathrm{ms}$~\cite{GW150914-CBC}. This $15$-ms window is determined by
the $10$-$\mathrm{ms}$ intersite propagation time plus an allowance for the uncertainty in identified
signal arrival times of weak signals. Candidate events are assigned a detection statistic value
 ranking their relative likelihood of being a gravitational-wave signal: 
the search uses an improved detection statistic compared to the first observing run~\cite{Nitz:2017}. 
The significance of a candidate event is calculated by comparing its detection statistic value to an
estimate of the background noise~\cite{Usman:2015kfa, GW150914-CBC, pycbc-github, Messick:2016aqy}. 
\EventName{} was detected with a network matched-filter signal-to-noise ratio (SNR) of \EventNetworkSNR.
At the detection statistic value assigned to \EventName{}, 
the false alarm rate is less than \EventFAR{} of coincident observing time.

The probability of astrophysical origin $P_{\mathrm{astro}}$ for a candidate event is 
found by comparing the candidate's detection statistic to a model described by 
the distributions and rates of both background
and signal events~\cite{GW150914-RATES, GW150914-RATES-SUPPLEMENT,Farr:2013yna}.
The background distribution is analysis dependent, being derived from the background samples used to calculate 
the false alarm rate. The signal distribution can depend on the mass distribution of the source systems; however, 
we find that different models of the binary black hole mass distribution (as described in Sec.~\ref{sec:discussion}) lead
to negligible differences in the resulting value of $P_{\mathrm{astro}}$. At the detection statistic
value of \EventName{}, the background rate in both matched filter
analyses is dwarfed by the signal rate, yielding $P_{\mathrm{astro}} > 1 - (3 \times 10^{-5})$.

An independent analysis that is not based on matched filtering, 
but instead looks for generic gravitational-wave bursts~\cite{GW150914-DETECTION,GW150914-BURST} and
selects events where the signal frequency rises over time~\cite{Klimenko:2015ypf},
also identified \EventName{}. 
This approach allows for signal deviations
from the waveform models used for matched filtering  
at the cost of a lower significance for signals that are
represented by the considered templates. 
This analysis reports a false alarm rate of \burstEventFAR{} for \EventName{}.

\section{Source properties}
\label{s:parameters}
The source parameters are inferred from a coherent Bayesian analysis of the data from both detectors~\cite{Veitch:2014wba,GW150914-PARAMESTIM}. 
As a cross-check, 
we use two independent model-waveform families. 
Both are tuned to numerical-relativity simulations of binary black holes with nonprecessing spins, 
and introduce precession effects through approximate prescriptions. 
One model includes inspiral spin precession using a single effective spin parameter $\chi_\mathrm{p}$~\cite{Hannam:2013oca,Schmidt:2014iyl,Khan:2015jqa}; 
the other includes the generic two-spin inspiral precession dynamics~\cite{Pan:2013rra,Taracchini:2013rva,Babak:2016tgq}. 
We refer to these as the effective-precession and full-precession models, respectively~\cite{GW150914-PRECESSING}. 
The two models yield consistent results.
Table~\ref{tab:parameters} shows selected source
parameters for \EventName{}; 
unless otherwise noted, we quote the median and symmetric $90\%$ credible interval for inferred quantities. 
The final mass (or equivalently the energy radiated), final spin and peak luminosity are computed using averages of fits to numerical-relativity results~\cite{Hofmann:2016yih,Jimenez-Forteza:2016oae,Healy:2016lce,T1600168,Keitel:2016krm}.
The parameter uncertainties include statistical and systematic errors
from averaging posterior probability distributions over the 
two waveform models, as well as calibration uncertainty~\cite{GW150914-PARAMESTIM} (and systematic uncertainty in the fit for peak luminosity). 
Statistical uncertainty dominates the overall uncertainty as a consequence of the moderate SNR.

\begin{table}
\caption{Source properties for \EventName{}: median values with $90\%$ credible intervals. 
We quote source-frame masses; to convert to the detector frame, multiply by 
$(1 + z)$~\cite{Krolak:1987ofj,Holz:2005df}.
The redshift assumes a flat cosmology with Hubble parameter $H_0 = 67.9~\mathrm{km\,s^{-1}\,Mpc^{-1}}$ and matter
density parameter $\Omega_\mathrm{m} = 0.3065$~\cite{Ade:2015xua}. 
\iftoggle{arXiv}{%
  More source properties are given in Table~\ref{tab:event-parameters} in the \textit{Supplemental Material}~\cite{Supp-Mat}.
}{%
  More source properties are given in Table~I of the \textit{Supplemental Material}~\cite{Supp-Mat}.
}%
}
\begin{ruledtabular}
\begin{tabular}{l l}
Primary black hole mass $m_1$ & $\MONESCOMPACTPerihelion\,\Msun$ \\
\rule{0pt}{3ex}%
Secondary black hole mass $m_2$ & $\MTWOSCOMPACTPerihelion\,\Msun$ \\
\rule{0pt}{3ex}%
Chirp mass $\mathcal{M}$ & $\MCSCOMPACTPerihelion\,\Msun$ \\
\rule{0pt}{3ex}%
Total mass $M$ & $\MTOTSCOMPACTPerihelion\,\Msun$ \\
\rule{0pt}{3ex}%
Final black hole mass $M_\mathrm{f}$ & $\MFINALSavgCOMPACTPerihelion\,\Msun$ \\
\rule{0pt}{3ex}%
Radiated energy $E_\mathrm{rad}$ & $\ERADCOMPACTPerihelion\,\Msun c^{2}$ \\
\rule{0pt}{3ex}%
Peak luminosity $\ell_\mathrm{peak} $& $\LPEAKCOMPACTPerihelion~\mathrm{erg\,s^{-1}}$ \\
\rule{0pt}{3ex}%
Effective inspiral spin parameter $\chi_\mathrm{eff}$ & $\CHIEFFCOMPACTPerihelion$ \\
\rule{0pt}{3ex}%
Final black hole spin $a_\mathrm{f}$ & $\SPINFINALinplaneCOMPACTPerihelion$ \\
\rule{0pt}{3ex}%
Luminosity distance $D_\mathrm{L}$ & $\DISTANCECOMPACTPerihelion~\mathrm{Mpc}$ \\
\rule{0pt}{3ex}%
Source redshift $z$ & $\REDSHIFTCOMPACTPerihelion$ \\
\end{tabular}
\end{ruledtabular}
\label{tab:parameters}
\end{table}

\begin{figure}
\centering
\includegraphics[width=0.98\columnwidth]{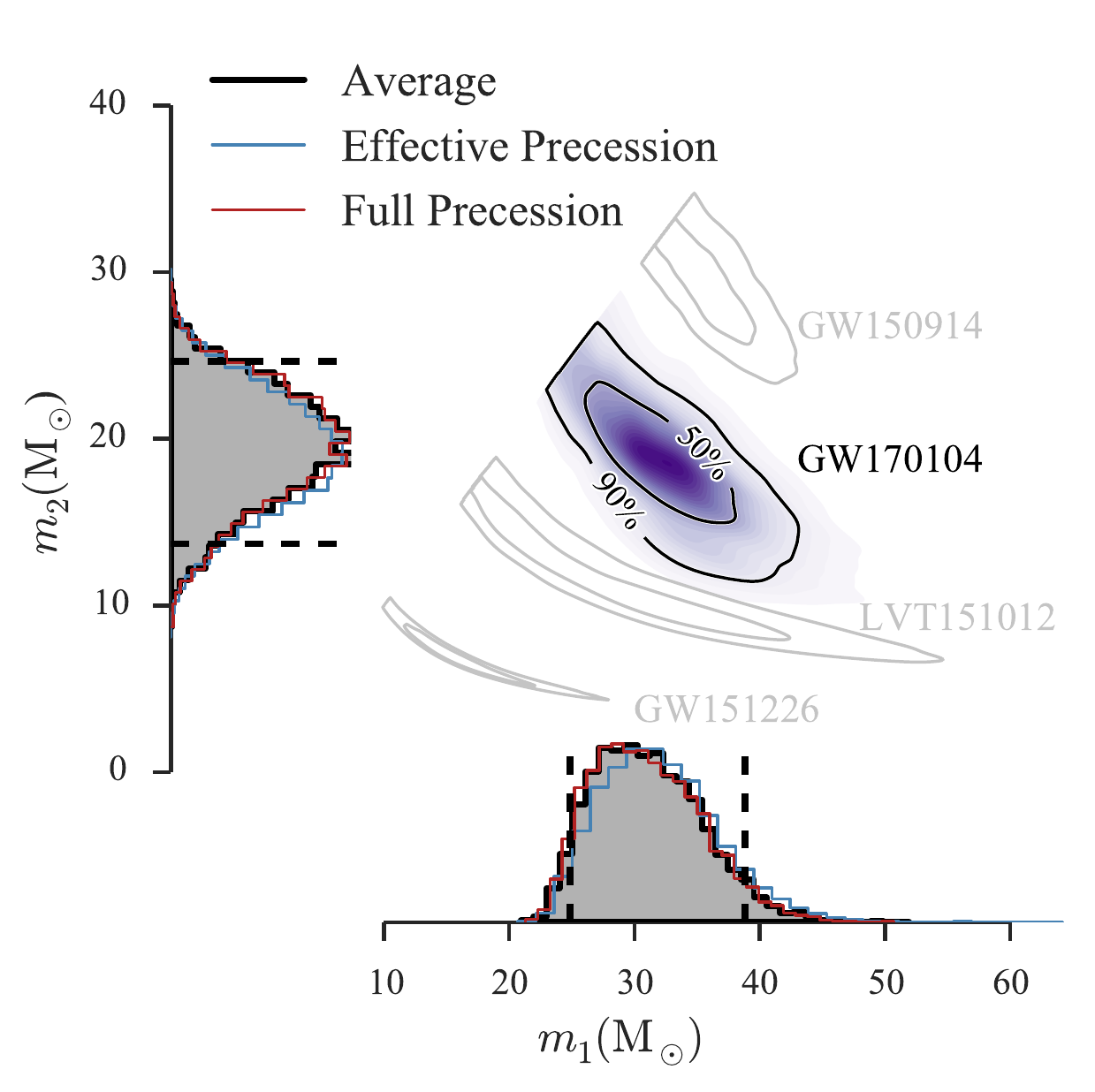}
\caption{Posterior probability density for the source-frame masses 
$m_1$ and $m_2$ (with $m_1 \geq m_2$). The 
one-dimensional distributions include the posteriors 
for the two 
waveform models, and their average (black). 
The dashed lines mark the $90\%$ credible interval for the average posterior.
The two-dimensional plot shows the contours of the $50\%$ and 
$90\%$ credible regions plotted over a color-coded posterior density function. 
For comparison, we also show the two-dimensional contours for the previous events~\cite{O1:BBH}.
}
\label{fig:mass}
\end{figure}

For binary coalescences, the gravitational-wave frequency evolution is primarily determined by the component masses.
For higher mass binaries, merger and ringdown dominate the signal, allowing good measurements of the total mass $M = m_1 + m_2$~\cite{Ajith:2009fz,Veitch:2015ela,Graff:2015bba,Haster:2015cnn,Ghosh:2015jra}.
For lower mass binaries, like GW151226~\cite{GW151226-DETECTION}, the inspiral is more important, providing precision measurements of the chirp mass $\mathcal{M} = (m_1 m_2)^{3/5}/M^{1/5}$~\cite{Finn:1992xs,Cutler:1994ys,Blanchet:1995ez,Vitale:2014mka}. 
The transition between the regimes depends upon the detectors' sensitivity, and \EventName{} sits between the two. 
The inferred component masses are shown in Fig.~\ref{fig:mass}. 
The form of the two-dimensional distribution is guided by the combination of constraints on $M$ and $\mathcal{M}$. 
The binary was composed of two black holes
with masses $m_{1} =\MONESCOMPACTPerihelion\,\Msun$ and $m_{2} =\MTWOSCOMPACTPerihelion\,\Msun$; 
these merged into a final black hole of mass $\MFINALSavgCOMPACTPerihelion\,\Msun$.  
This binary ranks second, behind GW150914's source~\cite{GW150914-PARAMESTIM,O1:BBH}, as the most massive stellar-mass binary black hole system observed to date.

\begin{figure}
 \centering
 \includegraphics[width=0.93\columnwidth]{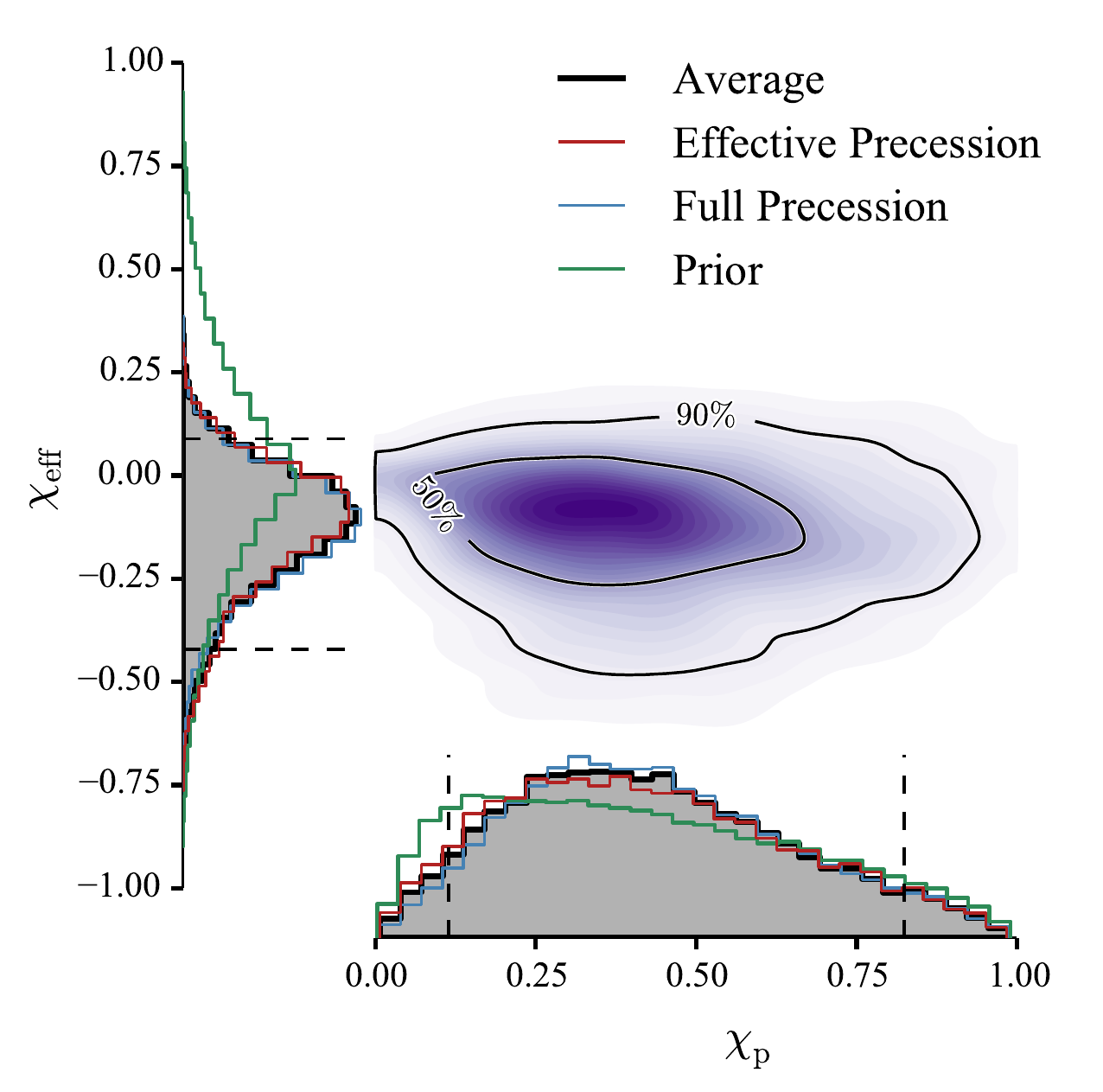} \\
 \includegraphics[width=0.88\columnwidth]{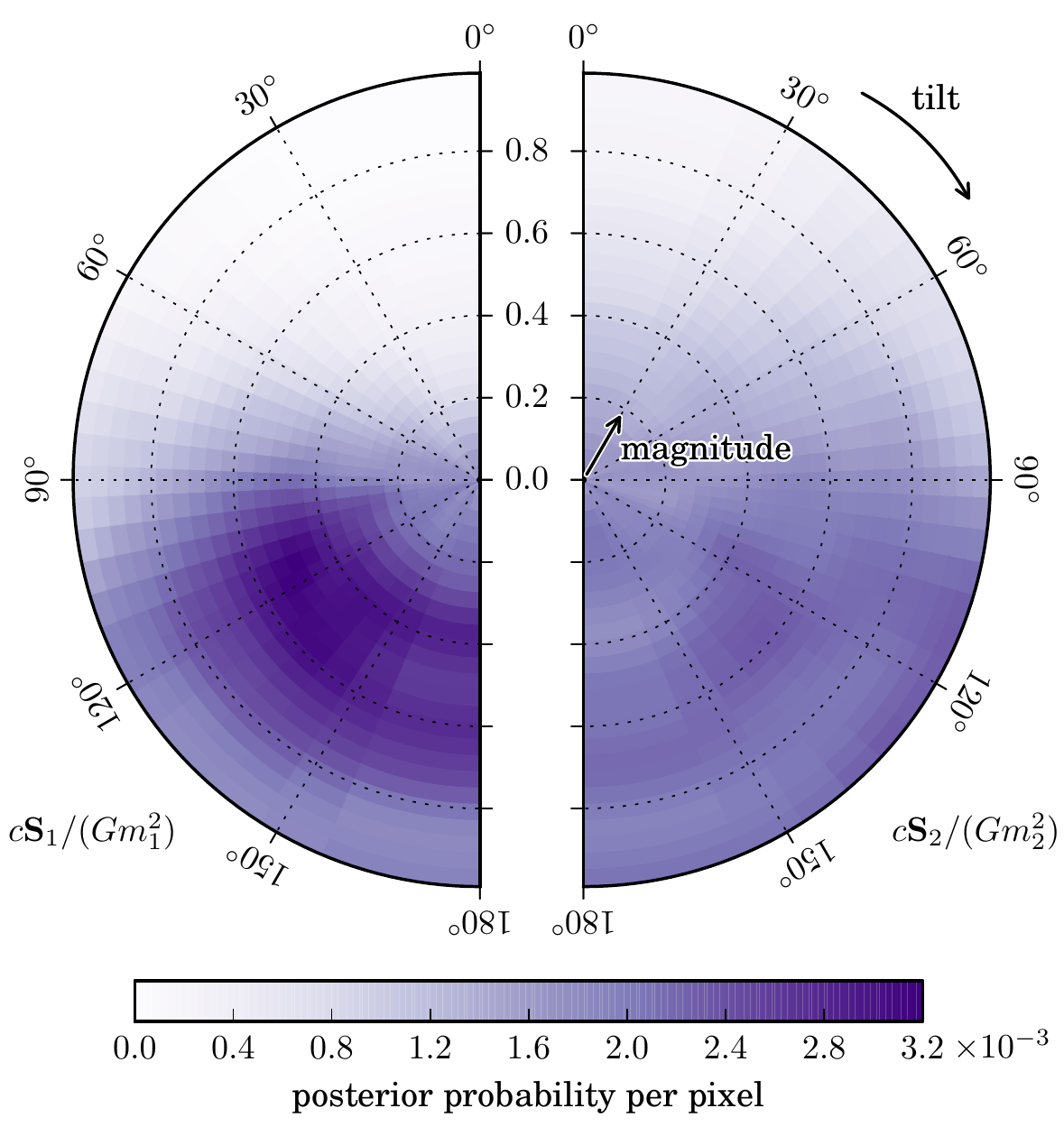}
 \caption{\textit{Top:} Posterior probability density for the effective inspiral and precession spin parameters,
$\chi_\mathrm{eff}$ and $\chi_\mathrm{p}$. 
The 
one-dimensional distributions show the posteriors 
for the two
waveform models, their average (black), and the prior distributions (green). 
The dashed lines mark the $90\%$ credible interval for the average posterior.
The two-dimensional plot shows the $50\%$ and 
$90\%$ credible regions plotted over the posterior density function. 
\textit{Bottom:} 
Posterior probabilities for the dimensionless component 
spins, $c\mathbf{S}_1/(Gm_1^2)$ and $c\mathbf{S}_2/(Gm_2^2)$, relative to the 
normal of the orbital plane $\mathbf{\hat{L}}$. 
The tilt angles are $0^\circ$ for spins aligned 
with the orbital angular momentum and $180^\circ$ for spins antialigned. 
The probabilities are marginalized over 
the azimuthal angles. 
The pixels have equal 
prior probability ($1.6\times10^{-3}$); they are spaced linearly in spin magnitudes and the 
cosine of the tilt angles.
Results are given at a gravitational-wave frequency of $20~\mathrm{Hz}$.
}
 \label{fig:spin}
\end{figure}

The black hole spins play a subdominant role in the orbital evolution of the binary, 
and are more difficult to determine. 
The orientations of the spins evolve due to precession~\cite{Apostolatos:1994mx,Blanchet:2013haa}, 
and we report results at a point in the inspiral corresponding to a gravitational-wave frequency of $20~\mathrm{Hz}$~\cite{GW150914-PARAMESTIM}. 
The effective inspiral spin parameter $\chi_\mathrm{eff} = (m_1 a_1 \cos\theta_{LS_1} + m_2 a_2 \cos\theta_{LS_2})/M$ 
is the most important spin combination for setting the properties of the inspiral~\cite{Damour:2001tu,Ajith:2009bn,Santamaria:2010yb} and remains important through to merger~\cite{Campanelli:2006uy,Reisswig:2009vc,Purrer:2013xma,Purrer:2015nkh,Vitale:2016avz}; 
it is approximately constant throughout the orbital evolution~\cite{Racine:2008qv,Gerosa:2015tea}. 
Here $\theta_{LS_i} = \cos^{-1}(\mathbf{\hat L} \cdot \mathbf{\hat S}_i)$ is the tilt angle between the spin $\mathbf{S}_i$ 
and the orbital angular momentum $\mathbf{L}$, 
which ranges from $0^\circ$ (spin aligned with orbital angular momentum) to $180^\circ$ (spin antialigned); 
$a_i = |c\mathbf{S}_i/Gm_i^2|$ is the (dimensionless) spin magnitude, which ranges from $0$ to $1$, and $i = 1$ for the primary black hole and $i = 2$ for the secondary. 
We use the Newtonian angular momentum for $\mathbf{L}$, such that it is normal to the orbital plane; the total orbital angular momentum differs from this because of post-Newtonian corrections. 
We infer that $\chi_\mathrm{eff} = \CHIEFFCOMPACTPerihelion$. 
Similarly to GW150914~\cite{GW150914-PARAMESTIM,GW150914-PRECESSING,O1:BBH}, $\chi_\mathrm{eff}$ is close to zero with a preference towards being negative: 
the probability that $\chi_\mathrm{eff} < 0$ is $\CHIEFFNEGPROBPerihelion$. 
Our measurements therefore disfavor a large total spin positively aligned with the orbital angular momentum, but do not exclude zero spins. 
 
The in-plane components of the spin control the amount of precession of the orbit~\cite{Apostolatos:1994mx}. 
This may be quantified by the effective precession spin parameter $\chi_\mathrm{p}$ which ranges from $0$ (no precession) to $1$ (maximal precession)~\cite{Schmidt:2014iyl}.
Figure~\ref{fig:spin} (top) shows the posterior probability density for $\chi_\mathrm{eff}$ and $\chi_\mathrm{p}$~\cite{Schmidt:2014iyl}. 
We gain some information on $\chi_\mathrm{eff}$, excluding large positive values,
but, as for previous events~\cite{GW150914-PARAMESTIM,GW151226-DETECTION,O1:BBH}, the $\chi_\mathrm{p}$ posterior is dominated by the prior
\iftoggle{arXiv}{%
  (see Appendix~\ref{s:pe-appendix} in the \textit{Supplemental Material}~\cite{Supp-Mat}). 
}{
  (see Sec.~III of the \textit{Supplemental Material}~\cite{Supp-Mat}).
}
No meaningful constraints can be placed on the magnitudes of the in-plane spin components and hence precession. 

The inferred component spin magnitudes and orientations are shown in Fig.~\ref{fig:spin} (bottom). 
The lack of constraints on the in-plane spin components means that we learn almost nothing about the spin magnitudes. 
The secondary's spin is less well constrained as the less massive component has a smaller impact on the signal. 
The probability that the tilt $\theta_{LS_i}$ is less than $45^\circ$ is $\TILTANGLEONEPROBPerihelion$ for the primary black hole and $\TILTANGLETWOPROBPerihelion$ for the secondary, 
whereas the prior probability is $0.15$ for each. 
Considering the two spins together, 
the probability that both tilt angles are less than $90^\circ$ is $\ALIGNEDPROBPerihelion$. 
Effectively all of the information comes from constraints on $\chi_\mathrm{eff}$ combined with the mass ratio (and our prior of isotropically distributed orientations and uniformly distributed magnitudes)~\cite{O1:BBH}. 

The source's luminosity distance $D_\mathrm{L}$ is inferred from the signal amplitude~\cite{Schutz:1986gp,GW150914-PARAMESTIM}. 
The amplitude is inversely proportional to the distance, but also depends upon the binary's inclination~\cite{Cutler:1994ys,Nissanke:2009kt,Rodriguez:2013oaa,Farr:2015lna}.
This degeneracy is a significant source of uncertainty~\cite{Ghosh:2015jra,Vitale:2016avz}. 
The inclination has a bimodal distribution with broad peaks for face-on and face-off orientations
\iftoggle{arXiv}{
  (see Fig.~\ref{fig:distance} if the \textit{Supplemental Material}~\cite{Supp-Mat}).
}{
  (see Fig.~4 of the \textit{Supplemental Material}~\cite{Supp-Mat}).
}
\EventName{}'s source is at $D_\mathrm{L} = \DISTANCECOMPACTPerihelion~\mathrm{Mpc}$, corresponding to a cosmological redshift of
$z = \REDSHIFTCOMPACTPerihelion$~\cite{Ade:2015xua}. 
While \EventName{}'s source has masses and spins comparable to GW150914's, it is most probably at a greater distance~\cite{GW150914-PARAMESTIM,O1:BBH}.

For GW150914, extensive studies were made to verify the accuracy of the model
waveforms for parameter estimation through comparisons with numerical-relativity
waveforms~\cite{GW150914-ACCURACY,GW150914-NRCOMP}. \EventName{} is a similar
system to GW150914 and, therefore, it is unlikely that there are any significant
biases in our results as a consequence of waveform modeling. The lower SNR of
\EventName{} makes additional effects not incorporated in the waveform models,
such as higher modes~\cite{Varma:2014jxa,Graff:2015bba,Bustillo:2015qty}, less
important. However, if the source is edge on or strongly precessing, there
could be significant biases in quantities including $\mathcal{M}$ and
$\chi_\mathrm{eff}$~\cite{GW150914-ACCURACY}. Comparison to 
numerical-relativity simulations of binary black holes with nonprecessing spins~\cite{GW150914-NRCOMP}, 
including those designed to replicate \EventName{}, produced results 
(and residuals) consistent with the model-waveform analysis.

\section{Waveform reconstructions}
\label{s:reconstructions}
\begin{figure}
\centering
\includegraphics[width=1\columnwidth]{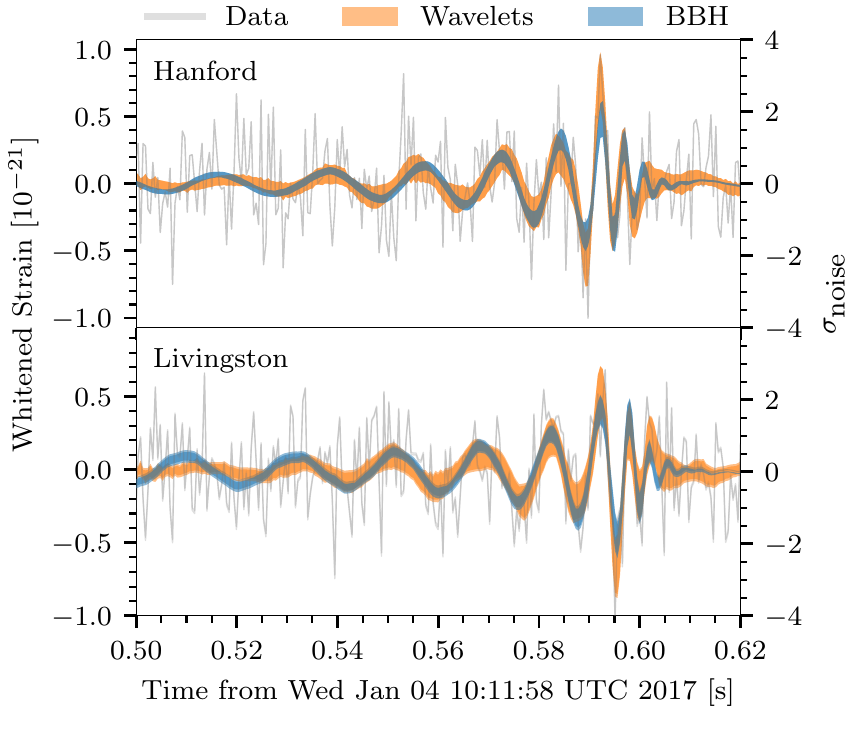}
\caption{Time-domain detector data (gray), and $90\%$ confidence intervals for
waveforms reconstructed from the morphology-independent wavelet analysis (orange) and
binary black hole (BBH) models from both waveform families (blue), whitened by 
each instrument's noise amplitude spectral density.  
The left ordinate axes are normalized such that the
amplitude of the whitened data and the physical strain are equal at
$200~\mathrm{Hz}$.  The right ordinate axes are in units of noise
standard deviations.  The width of the BBH region is dominated by the
uncertainty in the astrophysical parameters.}
\label{fig:wf_reconstructions}
\end{figure}
Consistency of \EventName{} with binary black hole waveform models can also be
explored through comparisons with a morphology-independent signal 
model~\cite{Cornish:2014kda}.  We choose to describe the signal as a
superposition of an arbitrary number of Morlet--Gabor wavelets, which models
an elliptically polarized, coherent signal in the detector
network.  Figure~\ref{fig:wf_reconstructions} plots whitened detector data at
the time of \EventName, together with waveforms drawn from the $90\%$ credible
region of the posterior distributions of the morphology-independent model
and the binary black hole waveform models used to infer the
source properties.  The signal appears in the two detectors with slightly
different amplitudes, and a relative phase shift of approximately $180^{\circ}$,
because of their different spatial orientations~\cite{GW150914-DETECTION}.  The wavelet- and template-based
reconstructions differ at early times because the wavelet
basis requires high-amplitude, well-localized signal energy to justify the
presence of additional wavelets, while the earlier portion of the
signal is inherently included in the binary black hole waveform model.  

The waveforms reconstructed from the morphology-independent model are consistent
with the characteristic inspiral--merger--ringdown structure.  The
overlap~\cite{Finn:1992xs} between the maximum-likelihood waveform of the binary
black hole model and the median waveform of the morphology-independent analysis
is $\burstOverlap\%$, consistent with expectations from Monte Carlo analysis of
binary black hole signals injected into detector data~\cite{GW150914-BURST}. 
We also use the morphology-independent analysis to search for residual
gravitational-wave energy after subtracting the maximum-likelihood binary black
hole signal from the measured strain data.  
There is an $83\%$ posterior probability in favor of Gaussian noise versus
residual coherent gravitational-wave energy which is not described by the
waveform model, implying that \EventName's source is a black hole binary.

\section{Binary black hole populations and merger rates}  
\label{sec:discussion}
The addition
of the first $11$ days of coincident observing time in the second observing run, 
and the detection of \EventName{}, 
leads to an improved estimate of the rate density of binary black hole mergers. 
We adopt two simple representative astrophysical population models: 
a distribution that is a power law in $m_1$ and uniform in $m_2$,  
$p\left( m_1, m_2 \right) \propto m_1^{-\alpha}/ \left(m_1 - 5 \,  \Msun\right)$ 
with $\alpha = 2.35$~\cite{Salpeter:1955it}, 
and a distribution uniform in the logarithm of each of the component 
masses~\cite{GW150914-RATES,O1:BBH}. 
In both cases, we impose $m_1, m_2 \geq 5 \, \Msun$ and 
$M \leq 100 \, \Msun$~\cite{GW150914-RATES}. 
Using the results from 
the first observing run as a prior, we obtain updated rates estimates 
of $\rateintrvcombimf{}$ for the power law, and $\rateintrcomblog{}$ 
for the uniform-in-log distribution~\cite{O1:BBH}. 
These combine search results from the 
two offline matched filter analyses, and marginalize over the calibration 
uncertainty~\cite{GW150914-RATES-SUPPLEMENT}. 
The range for the 
merger rate that brackets the two distributions, $\rateintrbracket$, 
is consistent with the range $9$--$240~\mathrm{Gpc^{-3}\,yr^{-1}}$ 
estimated from the first observing run~\cite{GW150914-RATES,O1:BBH}. 
Recalculating the rates directly after observing a new event can bias 
rate estimates, but this bias decreases with increasing event count 
and is negligible compared to other uncertainties on the intervals. 
While the median estimates have not changed appreciably, the overall 
tightening in the credible intervals is consistent with the additional 
observation time and the increment in the number of events with 
significant probability of being astrophysical from $3$ to $4$.

Following the first observing run, we performed a hierarchical analysis 
using the inferred masses of GW150914, LVT151012 and GW151226 to constrain 
the binary black hole mass distribution. 
We assumed the power-law 
population distribution described above, treating $\alpha$ as a parameter to 
be estimated, and 
found $\alpha = {2.5}^{+1.5}_{-1.6}$~\cite{O1:BBH}. 
With 
the addition of \EventName{}, 
\iftoggle{arXiv}{%
  $\alpha$ is estimated to be $\pmexpintr$ (see Appendix~\ref{s:pop-inf} in the \textit{Supplemental Material}~\cite{Supp-Mat}); 
}{
  $\alpha$ is estimated to be $\pmexpintr$ (see Sec.~IV of the \textit{Supplemental Material}~\cite{Supp-Mat});
}
the median is close to
the power-law exponent used to infer the (higher) merger rates. 

\section{Astrophysical implications}
\EventName{}'s source is a heavy stellar-mass binary black hole system. 
Such binaries are consistent with formation
through several different evolutionary pathways~\cite{GW150914-ASTRO}. 
Assuming black holes of stellar origin, there are two broad families of formation channels: dynamical and isolated binary evolution.  
Dynamical assembly of binaries is expected in dense stellar clusters~\cite{PortegiesZwart:1999nm,Rodriguez:2016avt,OLeary:2016ayz,Mapelli:2016vca,Banerjee:2016ths,Hurley:2016zkb,Askar:2016jwt,Park:2017zgj}. 
Dynamical influences are also important for binary coalescences near galactic nuclei~\cite{Stone:2016wzz,Bartos:2016dgn,Antonini:2016gqe}, 
and through interactions as part of a triple~\cite{Silsbee:2016djf,Antonini:2017ash}. 
Isolated binary evolution in galactic fields classically proceeds via a common envelope~\cite{Tutokov:1973,Voss:2003ep,Mennekens:2013dja,Belczynski:2016obo,Eldridge:2016ymr,Lipunov:2016cml,Woosley:2016nnw,Kruckow:2016tti,Stevenson:2017tfq}. 
Variants avoiding common-envelope evolution 
include (quasi-)chemically homogeneous evolution of massive tidally locked
binaries~\cite{Marchant:2016wow,Mandel:2015qlu,Eldridge:2016ymr}, 
or through stable mass transfer in Population I~\cite{Pavlovskii:2016edh,Heuvel:2017sfe} 
or Population III binaries~\cite{Hartwig:2016nde,Inayoshi:2017mrs}.

Stars lose mass throughout their lives; 
to leave a heavy black hole as a remnant they must avoid significant mass loss.  
Low-metallicity progenitors are believed to 
have weaker stellar winds and hence diminished mass loss~\cite{Vink:2008}. 
Given the mass of the primary black hole, the progenitors of \EventName{} likely formed in a 
lower metallicity environment $Z \lesssim 0.5 Z_{\odot}$~\cite{Belczynski:2009xy,Spera:2015vkd,Belczynski:2016obo,Rodriguez:2016kxx,GW150914-ASTRO}, 
but low mass loss may also have been possible at higher metallicity if the stars were strongly magnetized~\cite{Petit:2016uxn}. 

An alternative to the stellar-evolution channels would be binaries of primordial black holes~\cite{Carr:1974nx,Bird:2016dcv,Clesse:2016vqa,Carr:2016drx}. 
\EventName{}'s component masses lie in a range for which primordial black holes could contribute significantly to the dark matter 
content of the Universe, 
but merger rates in such scenarios are uncertain~\cite{Bird:2016dcv,Sasaki:2016jop}. 
The potential for existing electromagnetic observations to exclude primordial black holes of these masses is 
an active area of research~\cite{Ricotti:2007au,Clesse:2015wea,Clesse:2016vqa,Kashlinsky:2016sdv,Green:2016xgy,Ali-Haimoud:2016mbv,Mediavilla:2017bok,Carr:2017jsz}. 

Some of the formation models listed above predict merger rates on the order of $\sim1$--$10~\mathrm{Gpc}^{-3}\,\mathrm{yr}^{-1}$~\cite{Rodriguez:2016kxx,Rodriguez:2016avt,Mapelli:2016vca,Stone:2016wzz,Bartos:2016dgn,Antonini:2016gqe,Silsbee:2016djf,Antonini:2017ash,Mandel:2015qlu,Hartwig:2016nde}. 
Given that the rate intervals have now tightened and the lower bound 
(from the uniform-in-log distribution) is $\sim \rateintrbracketlow$, 
these channels may be insufficient to explain the full rate, 
but they could contribute to the total rate if there are multiple channels in operation. 
Future observations will improve the precision of the rate estimation, its redshift dependence, and our knowledge of the mass distribution, 
making it easier to constrain binary formation channels.

Gravitational-wave observations provide information about the
component spins through measurements of $\chi_\mathrm{eff}$, 
and these
measurements can potentially be used to distinguish different
formation channels. 
Dynamically assembled binaries (of both stellar
and primordial black holes) should have an isotropic distribution of
spin tilts, with equal probability for
positive and negative $\chi_\mathrm{eff}$, and a concentration around
zero~\cite{Rodriguez:2016vmx}. 
Isolated binary evolution typically predicts 
moderate ($\lesssim 45^\circ$) spin misalignments~\cite{Kalogera:1999tq}, 
since the effect of many astrophysical
processes, such as mass transfer~\cite{Bardeen:1975zz,King:2005mv} and
tides~\cite{Zahn:1977mi,Hut:1981}, is to align spins with the orbital
angular momentum. 
Black hole spins could become misaligned due to supernova explosions or torques during collapse. 
Large natal kicks are needed to produce negative $\chi_\mathrm{eff}$ 
by changing the orbital plane~\cite{Kalogera:1999tq,Rodriguez:2016vmx,OShaughnessy:2017eks}.
The
magnitude of these kicks is currently
uncertain~\cite{Martin:2009ny,Wong:2013vya,Janka:2013hfa,Repetto:2015kra,Mandel:2015eta,Repetto:2017gry}
and also influences the merger rate, with high kicks producing lower
merger rates in some population-synthesis
models~\cite{Voss:2003ep,Dominik:2013tma,Belczynski:2016obo,Rodriguez:2016kxx}. 
For binary neutron stars 
there is evidence that large tilts may be possible with small kicks~\cite{Spruit:1998sg,Farr:2011gs,Ferdman:2013xia,Kazeroni:2017fup}, 
and it is not yet understood if similar torques could occur for black holes~\cite{Janka:2013hfa,Miller:2014aaa,Fragos:2014cva,Fuller:2015}. 
The
absolute value of $\chi_\mathrm{eff}$ depends on the spin
magnitudes. 
Small values of
$|\chi_\mathrm{eff}|$ can arise because the spin magnitudes are low,
or because they are misaligned with the orbital angular momentum or
each other. 
The spin magnitudes for binary black holes are currently uncertain, 
but GW151226 demonstrated that
they can be $\gtrsim
\SPINMAXMINBoxing$~\cite{GW151226-DETECTION}, and high-mass x-ray
binary measurements indicate that the distribution of black hole spins could extend to larger
magnitudes~\cite{Miller:2014aaa}. 
For \EventName{}, we infer $\chi_\mathrm{eff} = \CHIEFFCOMPACTPerihelion$. 
This includes the possibility of negative $\chi_\mathrm{eff}$, which would indicate spin--orbit misalignment of at least one component.
It also excludes large positive values, and thus could argue
against its source forming through chemically homogeneous evolution, 
since large aligned spins ($a_{i} \gtrsim 0.4$) would be expected
assuming the complete collapse of the progenitor stars~\cite{Marchant:2016wow}. 
The inferred range is consistent with dynamical assembly 
and 
isolated binary evolution provided that the positive orbit-aligned spin is small (whether due to low spins or misalignment)~\cite{Zaldarriaga:2017qkw,Hotokezaka:2017esv,Gerosa:2013laa,Rodriguez:2016vmx}. 
Current gravitational-wave measurements cluster around
$\chi_\mathrm{eff} \sim 0$ ($|\chi_\mathrm{eff}| <
\CHIEFFMAGLIMITBoxing$ at the $90\%$ credible level for all events;
\iftoggle{arXiv}{%
  see Fig.~\ref{fig:chi-all} in the \textit{Supplemental Material}~\cite{Supp-Mat})~\cite{O1:BBH}.
}{
  see Fig.~5 of the \textit{Supplemental Material}~\cite{Supp-Mat})~\cite{O1:BBH}.
}
Assuming that binary black hole spins are not typically small ($\lesssim 0.2$), our
observations hint towards the astrophysical population favoring
a distribution of misaligned spins rather than near orbit-aligned spins~\cite{Farr:2017}; further
detections will test if this is the case, 
and enable us to distinguish different spin magnitude and orientation distributions~\cite{Gerosa:2014kta,Vitale:2015tea,Gerosa:2017kvu,Fishbach:2017dwv,Stevenson:2017dlk,Talbot:2017yur}.

\section{Tests of general relativity}
\label{s:grtests}
To check the consistency of the observed signals with the predictions of GR for binary black holes in quasicircular orbit, we employ a 
phenomenological approach that probes how gravitational-wave generation or
propagation could be modified in an alternative theory of gravity.  Testing for
these characteristic modifications in the waveform can quantify the degree
to which departures from GR can be tolerated given the data.  First,   
we consider the possibility of a modified gravitational-wave dispersion relation, 
and place bounds on the magnitude of potential deviations from GR. 
Second, we perform null tests to quantify generic deviations from GR: without assuming a specific alternative theory of gravity, we verify if the detected signal is compatible with GR. 
For these tests we use the three confident detections (GW150914, GW151226 and \EventName{}); 
we do not use the marginal event LVT151012, as its low SNR means that it contributes insignificantly to all the tests~\cite{O1:BBH}.

\subsection{Modified dispersion}

In GR, gravitational waves are nondispersive.
We consider a modified
dispersion relation of the form $E^2=p^2 c^2+A p^{\alpha} c^\alpha$,
$\alpha\geq0$, that leads to dephasing of the waves relative
to the phase evolution in GR. Here $E$ and $p$ are the energy and
momentum of gravitational radiation, and $A$ is the amplitude of
the dispersion~\cite{Mirshekari:2011yq,Yunes:2016jcc}.  Modifications to the dispersion
relation can arise in theories that include violations of local Lorentz
invariance~\cite{Mattingly:2005re}.  Lorentz invariance is a cornerstone of
modern physics but its violation is expected in certain quantum gravity
frameworks~\cite{Will:2014kxa,Mattingly:2005re}. Several modified theories 
of gravity predict specific values of $\alpha$,
including massive-graviton theories ($\alpha=0$, $\lora>0$)~\cite{Will:2014kxa},
multifractal spacetime~\cite{Calcagni:2009kc} ($\alpha=2.5$), doubly special
relativity~\cite{AmelinoCamelia:2002wr} ($\alpha=3$), and Ho{\v r}ava--Lifshitz
\cite{Horava:2009uw} and extra-dimensional~\cite{Sefiedgar:2010we} theories 
($\alpha=4$). For our analysis, we assume that the only effect of these alternative theories 
is to modify the dispersion relation.

\begin{figure}
 \centering
 \includegraphics[width=1\columnwidth]{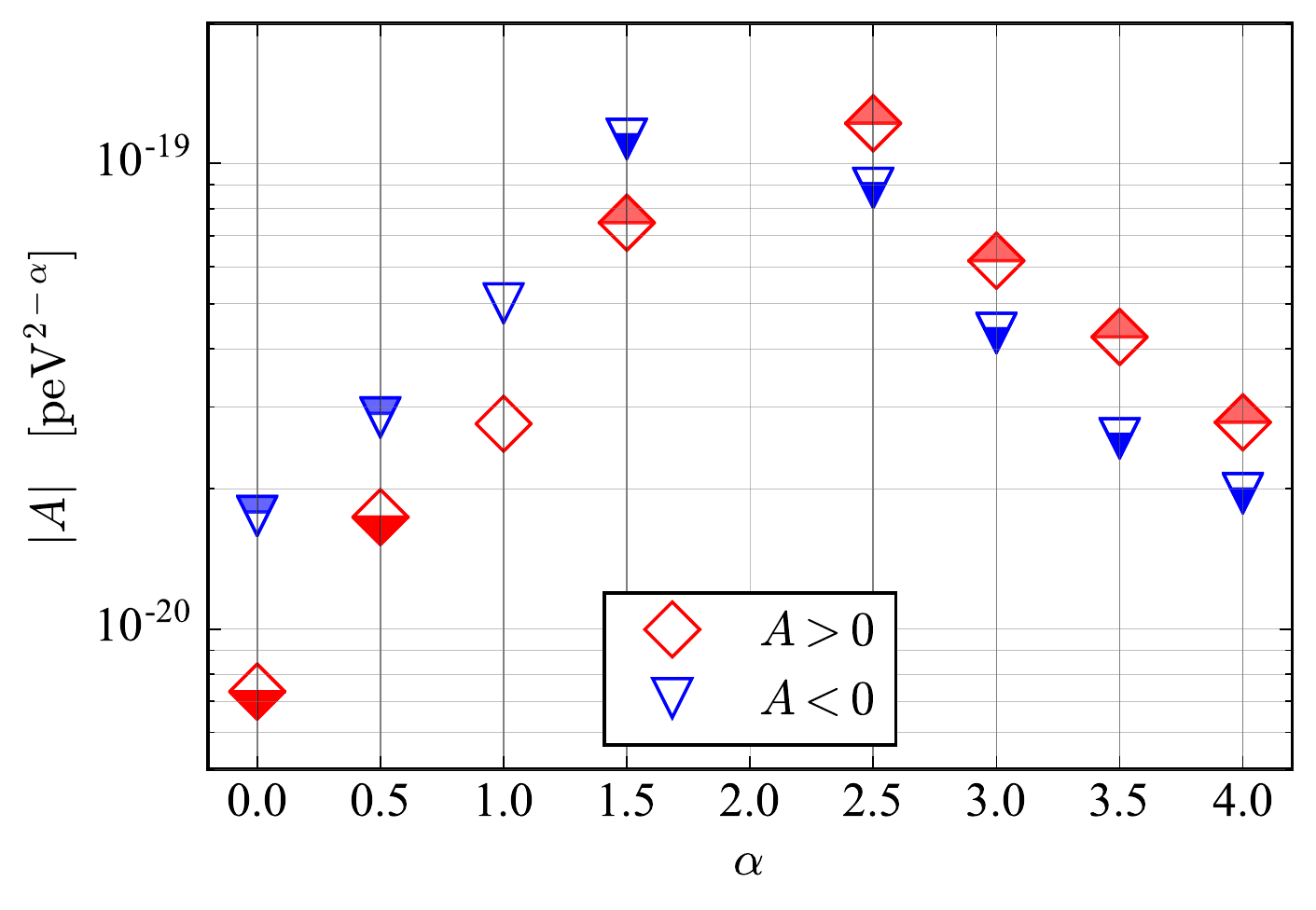}
 \caption{$90\%$ credible upper bounds on $|A|$, the magnitude of dispersion,
   obtained combining the posteriors of \EventName{}
 with those of GW150914 and GW151226. 
 We use picoelectronvolts as a convenient unit because the corresponding frequency scale is around where \EventName{} has greatest amplitude ($1~\mathrm{peV} \simeq h \times 250~\mathrm{Hz}$, where $h$ is the Planck constant).   
 General relativity corresponds to $A=0.$ Markers filled at the top (bottom)
 correspond to values of $|A|$ and $\alpha$ for which gravitational waves travel with superluminal
 (subluminal) speed.}
 \label{fig:tgr_liv}
\end{figure}

To leading order in $A E^{\alpha-2}$, the group velocity of gravitational waves
is modified as $v_g/c = 1 + (\alpha -1) A E^{\alpha-2} /2$~\cite{Yunes:2016jcc};  both superluminal
and subluminal propagation velocities are possible, depending on the sign of $A$
and the value of $\alpha$.  A change in the dispersion relation leads to an
extra term $\delta\Psi(A,\alpha)$ in the evolution
of the gravitational-wave phase~\cite{Mirshekari:2011yq}. We introduce such a term in the effective-precession
waveform model~\cite{Hannam:2013oca} to constrain dispersion for various values of $\alpha.$ 
To this end, we assume flat priors on $A$.  
In Fig.~\ref{fig:tgr_liv} we show $90\%$ credible upper bounds on $|A|$ derived 
from the three confident detections. 
We do not show results for $\alpha=2$ since in 
this case the modification of the gravitational-wave phase is degenerate with 
the arrival time of the signal. 

There exist constraints on Lorentz invariance violating dispersion relations from other observational sectors (e.g.,
photon or neutrino observations) for certain values of $\alpha$, and  our results are 
weaker by several orders of magnitude.  However, there are frameworks in which Lorentz invariance is
only broken in one sector~\cite{Seifert:2009gi,Altschul:2009ae}, implying that each sector
provides complementary information on potential modifications to GR.  Our results are the first
bounds derived from gravitational-wave observations, and the first tests of superluminal propagation
in the gravitational sector.
 
The result for $A>0$ and $\alpha=0$ can be reparametrized to derive a lower
bound on the graviton Compton wavelength $\lambda_{g}$, assuming that gravitons disperse in
vacuum in the same way as massive particles~\cite{Will:1997bb,
GW150914-TESTOFGR, O1:BBH}. In this case, no violation of Lorentz invariance
is assumed. Using a flat prior for the graviton mass, we obtain
$\lambda_{g}>\LAMBDAMGPerihelion~\mathrm{km}$, which improves on the
bound of $\LAMBDAMGGWFIRSTEVENT~\mathrm{km}$ from previous gravitational-wave
observations~\cite{GW150914-TESTOFGR,O1:BBH}.  The combined bound using the 
three confident detections is $\lambda_{g}>\LAMBDAMGCumulative~\mathrm{km}$, 
or for the graviton mass $m_g\leq\MASSMGCumulative~\mathrm{eV}/c^2$.  

\subsection{Null tests}

In the post-Newtonian approximation, the gravitational-wave phase in the Fourier
domain is a series expansion in powers of frequency, the expansion coefficients
being functions of the source
parameters~\cite{Blanchet:1995ez,Blanchet:2004ek,Blanchet:2013haa}. In the
effective-precession model, waveforms from numerical-relativity simulations are
also modeled using an expansion of the phase in terms of the Fourier frequency.
To verify if the detected signal is consistent with GR, we allow the expansion
coefficients to deviate in turn from their nominal GR value and we obtain a
posterior distribution for the difference between the measured and GR
values~\cite{Blanchet:1994ex,Blanchet:1995fg,Arun:2006hn,Mishra:2010tp,Yunes:2009ke,Li:2011cg}.
We find no significant deviation from the predictions of
GR~\cite{GW150914-TESTOFGR,O1:BBH}.  Combined bounds for
\EventName{} and the two confident detections from the first observing run~\cite{O1:BBH}
do not significantly improve the bounds on the waveform phase coefficients.  

Finally, we investigate whether the merger--ringdown portion of the
detected signal is consistent with the inspiral 
part~\cite{GW150914-TESTOFGR,Ghosh:2016qgn,Ghosh:2017gfp}.
The two parts are divided at $\IMRTRANSITIONFREQPerihelion{}~\mathrm{Hz}$, a frequency close to the median inferred (detector-frame) innermost-stable-circular-orbit frequency of the remnant Kerr black hole. 
For each part, we infer the component masses and spins, and calculate from these the
final mass and spin using fits from numerical
relativity, as in Sec.~\ref{s:parameters}~\cite{Hofmann:2016yih,Jimenez-Forteza:2016oae,Healy:2016lce,T1600168}.
We then calculate a two-dimensional posterior distribution for the fractional
difference between final mass and spin calculated separately from the two
parts~\cite{GW150914-TESTOFGR,Ghosh:2017gfp}.  The expected GR value (no
difference in the final mass and spin estimates) lies close to the peak of the
posterior distribution, well within the $90\%$ credible region.  When combined
with the posteriors from GW150914, the width of the credible intervals decreases
by a factor of $\sim 1.5$, providing a better constraint on potential deviations from GR.

In conclusion, in agreement with the predictions of GR, none of the tests we performed indicate a statistically
significant departure from the coalescence of Kerr black holes in a quasicircular orbit.

\section{Conclusions}
\label{s:outlook}
Advanced LIGO began its second observing run on November 30, 2016, and on January 4, 2017
the LIGO-Hanford and LIGO-Livingston detectors registered a highly significant gravitational-wave 
signal \EventName{} from the coalescence of two stellar-mass black holes.  
\EventName{} joins two other high-significance events~\cite{GW150914-DETECTION,GW151226-DETECTION} and a 
marginal candidate~\cite{GW150914-CBC} from Advanced LIGO's first observing run~\cite{O1:BBH}.  
This new detection is entirely consistent 
with the astrophysical rates inferred from the previous run. 
The source is a heavy binary black hole system, similar to that of GW150914.
Spin configurations with both component spins aligned with the orbital angular momentum 
are disfavored (but not excluded);  
we do not significantly constrain the component black holes' spin magnitudes. 
The observing run will continue until mid 2017. 
Expanding the catalog of binary black holes will provide further insight into their formation and evolution, 
and allow for tighter constraints on potential modifications to GR.

Further details of the analysis and the results are given in the
\iftoggle{arXiv}{%
  \textit{Supplemental Material}~\cite{Supp-Mat}.
}{
  \textit{Supplemental Material}~\cite{Supp-Mat}.
}
Data for 
this event are available at the LIGO Open Science Center~\cite{LOSC:GW170104}.

\acknowledgments 
The authors gratefully acknowledge the support of the United States
National Science Foundation (NSF) for the construction and operation of the
LIGO Laboratory and Advanced LIGO as well as the Science and Technology Facilities Council (STFC) of the
United Kingdom, the Max-Planck-Society (MPS), and the State of
Niedersachsen/Germany for support of the construction of Advanced LIGO 
and construction and operation of the GEO\,600 detector. 
Additional support for Advanced LIGO was provided by the Australian Research Council.
The authors gratefully acknowledge the Italian Istituto Nazionale di Fisica Nucleare (INFN),  
the French Centre National de la Recherche Scientifique (CNRS) and
the Foundation for Fundamental Research on Matter supported by the Netherlands Organisation for Scientific Research, 
for the construction and operation of the Virgo detector
and the creation and support  of the EGO consortium. 
The authors also gratefully acknowledge research support from these agencies as well as by 
the Council of Scientific and Industrial Research of India, 
Department of Science and Technology, India,
Science \& Engineering Research Board (SERB), India,
Ministry of Human Resource Development, India,
the Spanish Ministerio de Econom\'ia y Competitividad,
the  Vicepresid\`encia i Conselleria d'Innovaci\'o, Recerca i Turisme and the Conselleria d'Educaci\'o i Universitat del Govern de les Illes Balears,
the National Science Centre of Poland,
the European Commission,
the Royal Society, 
the Scottish Funding Council, 
the Scottish Universities Physics Alliance, 
the Hungarian Scientific Research Fund (OTKA),
the Lyon Institute of Origins (LIO),
the National Research Foundation of Korea,
Industry Canada and the Province of Ontario through the Ministry of Economic Development and Innovation, 
the Natural Science and Engineering Research Council Canada,
Canadian Institute for Advanced Research,
the Brazilian Ministry of Science, Technology, and Innovation,
International Center for Theoretical Physics South American Institute for Fundamental Research (ICTP-SAIFR), 
Russian Foundation for Basic Research,
the Leverhulme Trust, 
the Research Corporation, 
Ministry of Science and Technology (MOST), Taiwan
and
the Kavli Foundation.
The authors gratefully acknowledge the support of the NSF, STFC, MPS, INFN, CNRS and the
State of Niedersachsen/Germany for provision of computational resources. This article has been assigned the document
number LIGO-P170104.

\iftoggle{arXiv}{%
  \clearpage
  \section*{Supplemental Material}
  \label{s:supplement}
  
  \appendix
  \section{Noise performance of the detectors}
\label{s:detectors_extra}
\begin{figure*}[!bth]
\vskip-10pt
 \centering
\includegraphics[width=0.99\textwidth]{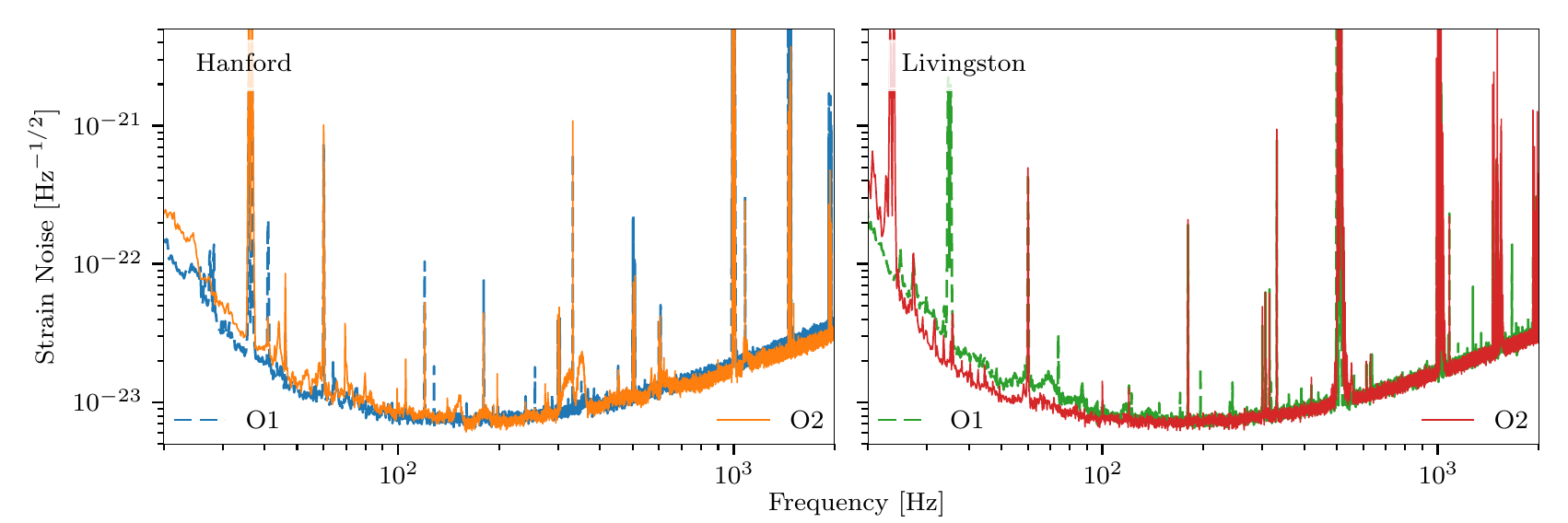}
\vskip-10pt
\caption{Comparison of typical noise amplitude spectra of the LIGO detectors in the first observing run (O1) and the early stages of the second observing run (O2).
The noise is expressed in terms of equivalent gravitational-wave strain amplitude.
Some narrow features are calibration lines ($22$--$24~\mathrm{Hz}$ for L1, $35$--$38~\mathrm{Hz}$ for H1, $330~\mathrm{Hz}$ and $1080~\mathrm{Hz}$ for both), suspension fibers' resonances ($500~\mathrm{Hz}$ and harmonics) and $60~\mathrm{Hz}$ power line harmonics.}
 \label{fig:noiseASDO1O2}
\end{figure*}
Figure~\ref{fig:noiseASDO1O2} shows a comparison of typical strain noise
amplitude spectra during the first observing run and early in the second for both of the LIGO detectors~\cite{G1500623}.
For the Hanford detector, shot-noise limited performance was improved above about $500~\mathrm{Hz}$ by increasing the laser power. There are new broad mechanical resonance features (e.g., at $\sim 150~\mathrm{Hz}$, $320~\mathrm{Hz}$ and $350~\mathrm{Hz}$) due to increased beam pointing jitter from the laser, as well as  the coupling of the jitter to the detector's gravitational-wave channel that is larger than in the Livingston detector. The increase in the noise between $40~\mathrm{Hz}$ and $100~\mathrm{Hz}$ is currently under investigation.
For the Livingston detector, significant reduction in the noise between $25~\mathrm{Hz}$ and $100~\mathrm{Hz}$ was achieved mainly by the reduction of the scattered light that re-enters the interferometer.

To date, the network duty factor of the LIGO detectors in the second observing run is about
$51\%$ while it was about $43\%$ in the first observing run.
The improvement came from better seismic isolation at Hanford, and
fine tuning of the control of the optics at Livingston.

  \begin{figure*}
 \centering
 \includegraphics[width=0.98\columnwidth]{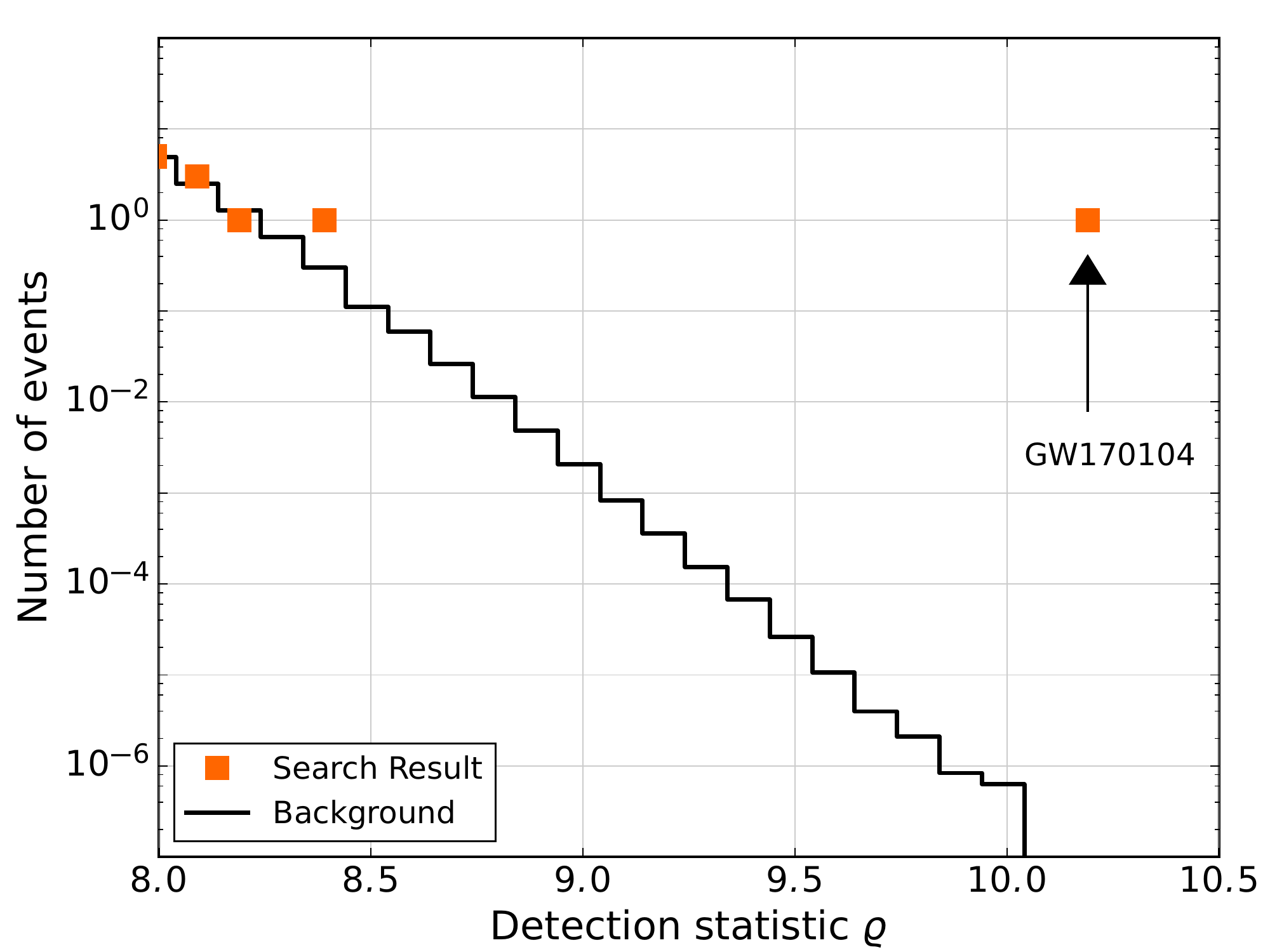}
 \includegraphics[width=0.98\columnwidth]{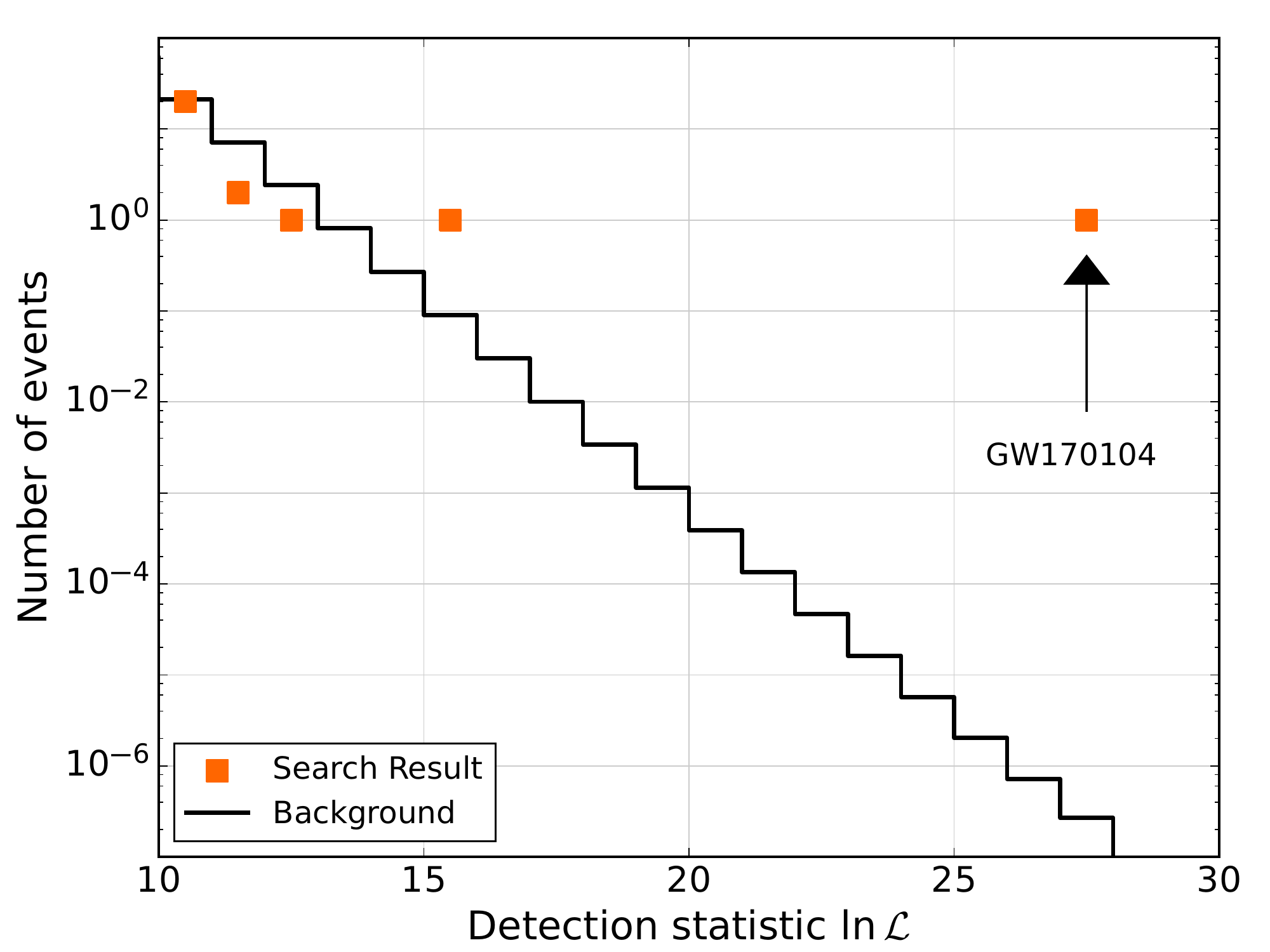}
 \caption{
\textit{Left:}  Search results from the binary coalescence search described  
in~\cite{Usman:2015kfa,Nitz:2017,pycbc-github}. 
The histogram shows the number of candidate events (orange markers) in the
\OTwoChunkThreeLength{} days of coincident data
and the expected background (black lines) as a function of
the search detection statistic. The reweighted SNR detection statistic $\varrho$ is defined in~\cite{Nitz:2017}. 
\EventName{} has a larger detection
statistic value than all of the background events in this period.
At the detection statistic value assigned to \EventName{},
the search's false alarm rate is less than \EventFAR{}
of coincident observing time. No other significant candidate events are
observed in this time interval. \textit{Right:} Search results from an 
independently-implemented analysis~\cite{Messick:2016aqy}, 
where the detection statistic $\ln\mathcal{L}$ is an approximate log likelihood ratio
statistic that is an extension of~\cite{Cannon:2015gha}. The two search algorithms give consistent results.
}
 \label{fig:bkg}
\end{figure*}

\section{Searches}
\label{s:searches_sup}

The significance of a candidate event is calculated by comparing its 
detection statistic value to an estimate of
the background noise~\cite{Usman:2015kfa, GW150914-CBC, pycbc-github, Messick:2016aqy, Nitz:2017}.
Figure~\ref{fig:bkg} shows the background and candidate events from the offline
searches for compact binary coalescences obtained from
\OTwoChunkThreeLength{} days of coincident data.
At the detection statistic value assigned to GW170104, the false alarm rate 
is less than \EventFAR{} of coincident observing time.

  \section{Parameter inference}
\label{s:pe-appendix}

The source properties are  
estimated by exploring the parameter space with stochastic sampling algorithms~\cite{Veitch:2014wba}. 
Calculating the posterior probability requires the likelihood of the data given a set of parameters, and the parameters' prior probabilities. 
The likelihood is determined from a noise-weighted inner product between the data and a template waveform~\cite{Cutler:1994ys}. 
Possible calibration error is incorporated using a frequency-dependent spline model for each detector~\cite{SplineCalMarg-T1400682}. The analysis follows the approach used for previous signals~\cite{GW150914-PARAMESTIM,GW150914-PRECESSING,O1:BBH}.

A preliminary analysis was performed to provide a medium-latency source localization~\cite{GCN20385}. This analysis used an initial calibration of the data and assumed a (conservative) one-sigma calibration uncertainty of $10\%$ in amplitude and $10^\circ$ in phase for both detectors, a reduced-order quadrature model of the effective-precession waveform~\cite{Smith:2016qas,Husa:2015iqa,Khan:2015jqa,Hannam:2013oca} (the most computationally expedient model), and a power spectral density calculated using a parametrized model of the detector noise~\cite{Cornish:2014kda,Littenberg:2014oda}. A stretch of $4~\mathrm{s}$ of data, centered on the event, was analysed across a frequency range of $20$--$1024~\mathrm{Hz}$. We assumed uninformative prior probabilities~\cite{GW150914-PARAMESTIM,O1:BBH}; 
technical restrictions of the reduced-order quadrature required us to limit spin magnitudes to $< 0.8$ and impose cuts on the masses (as measured in the detector frame) such that $m_{1,2}^\mathrm{det} \in [5.5, 160]\,\Msun$, $\mathcal{M}^\mathrm{det} \in [12.3, 45.0]\,\Msun$ and mass ratio $q = m_2/m_1 \geq 1/8$. The bounds of the mass prior do not affect the posterior, but the spin distributions were truncated. The source position is not strongly coupled to the spin distribution, and so should not have been biased by 
these limits~\cite{Singer:2015ema,Farr:2015lna}. 

The final analysis used an updated calibration of the data, with one-sigma uncertainties of $3.8\%$ in amplitude and $2.2^\circ$ in phase for Hanford, and $3.8\%$ and $1.9^\circ$ for Livingston, and two waveform models, 
the effective-precession model~\cite{Husa:2015iqa,Khan:2015jqa,Hannam:2013oca}
and the full-precession model~\cite{Pan:2013rra,Taracchini:2013rva,Babak:2016tgq}.
The spin priors were extended up to $0.99$. 
As a consequence of the computational cost of the full-precession model, we approximate the likelihood by marginalising over the time and phase at coalescence as if the waveform contained only the dominant $(2,\pm2)$ harmonics~\cite{Veitch:2014wba}. 
This marginalisation is not exact for precessing models, but should not significantly affect signals with binary inclinations that are nearly face on or face off~\cite{GW150914-PRECESSING}. 
Comparisons with preliminary results from an investigation using the full-precession waveform without marginalisation confirm that this approximation does not impact results. 
The two waveform models produce broadly consistent parameter estimates, so the overall results are constructed by averaging the two distributions. 
As a proxy for the theoretical error from waveform modeling, we use the difference between the results from the two approximants~\cite{GW150914-PARAMESTIM}. 
A detailed summary of results is given in Table~\ref{tab:event-parameters}, and the final sky localization is shown in Fig.~\ref{fig:sky}.

\begin{table*}
\caption{Parameters describing \EventName{}. We report the median value with symmetric (equal-tailed) $90\%$ credible interval, and selected $90\%$ credible bounds.
Results are given for effective- and full-precession waveform models; the overall 
results average the posteriors for the two models. 
The overall results include
a proxy for the $90\%$ range of systematic error estimated from the
variance between models.
More details of the parameters, and the imprint they leave on the 
signal, are explained in \cite{GW150914-PARAMESTIM}.
The optimal SNR is the noise-weighted inner product of the waveform template with itself, whereas the matched-filter SNR is the inner product of the template with the data.
}
\begin{ruledtabular}
\begin{tabular}{l c c c}
 & Effective precession & Full precession & Overall \\
\hline
\hline
Detector-frame & & & \\
\quad{}Total mass $M^\mathrm{det}/\Msun$ & \imrppMTOTobsPerihelion & \seobnrMTOTobsPerihelion & \MTOTobsPerihelion \\
\quad{}Chirp mass $\mathcal{M}^\mathrm{det}/\Msun$ & \imrppMCobsPerihelion & \seobnrMCobsPerihelion & \MCobsPerihelion \\
\quad{}Primary mass $m_1^\mathrm{det}/\Msun$ & \imrppMONEobsPerihelion & \seobnrMONEobsPerihelion & \MONEobsPerihelion \\
\quad{}Secondary mass $m_2^\mathrm{det}/\Msun$ & \imrppMTWOobsPerihelion & \seobnrMTWOobsPerihelion & \MTWOobsPerihelion \\
\quad{}Final mass $M_\mathrm{f}^\mathrm{det}/\Msun$ & \imrppMFINALobsavgPerihelion & \seobnrMFINALobsavgPerihelion & \MFINALobsavgPerihelion \\
\rule{0pt}{3ex}%
Source-frame & & & \\
\quad{}Total mass $M/\Msun$ & \imrppMTOTSPerihelion & \seobnrMTOTSPerihelion & \MTOTSPerihelion \\
\quad{}Chirp mass $\mathcal{M}/\Msun$ & \imrppMCSPerihelion & \seobnrMCSPerihelion & \MCSPerihelion \\
\quad{}Primary mass $m_1/\Msun$ & \imrppMONESPerihelion & \seobnrMONESPerihelion & \MONESPerihelion \\
\quad{}Secondary mass $m_2/\Msun$ & \imrppMTWOSPerihelion & \seobnrMTWOSPerihelion & \MTWOSPerihelion \\
\quad{}Final mass $M_\mathrm{f}/\Msun$ & \imrppMFINALSavgPerihelion & \seobnrMFINALSavgPerihelion & \MFINALSavgPerihelion \\
\quad{}Energy radiated $E_\mathrm{rad}/(\Msun c^2)$ & \imrppERADPerihelion & \seobnrERADPerihelion & \ERADPerihelion \\
\rule{0pt}{3ex}%
Mass ratio $q$ & \imrppMASSRATIOPerihelion & \seobnrMASSRATIOPerihelion & \MASSRATIOPerihelion \\
\rule{0pt}{3ex}%
Effective inspiral spin parameter $\chi_\mathrm{eff}$ & \imrppCHIEFFPerihelion & \seobnrCHIEFFPerihelion & \CHIEFFPerihelion \\
Effective precession spin parameter $\chi_\mathrm{p}$ & \imrppCHIPPerihelion & \seobnrCHIPPerihelion & \CHIPPerihelion \\
Dimensionless primary spin magnitude $a_1$ & \imrppSPINONEPerihelion & \seobnrSPINONEPerihelion & \SPINONEPerihelion \\
Dimensionless secondary spin magnitude $a_2$ & \imrppSPINTWOPerihelion & \seobnrSPINTWOPerihelion & \SPINTWOPerihelion \\
Final spin $a_\mathrm{f}$ & \imrppSPINFINALinplanePerihelion & \seobnrSPINFINALinplanePerihelion & \SPINFINALinplanePerihelion \\
\rule{0pt}{3ex}%
Luminosity distance $D_\mathrm{L}/\mathrm{Mpc}$ & \imrppDISTANCEPerihelion & \seobnrDISTANCEPerihelion & \DISTANCEPerihelion \\
Source redshift $z$ & \imrppREDSHIFTPerihelion & \seobnrREDSHIFTPerihelion & \REDSHIFTPerihelion \\
\rule{0pt}{3ex}%
Upper bound & & & \\
\quad{}Effective inspiral spin parameter $\chi_\mathrm{eff}$ & \imrppCHIEFFLIMITPerihelion & \seobnrCHIEFFLIMITPerihelion & \CHIEFFLIMITSYSPerihelion \\
\quad{}Effective precession spin parameter $\chi_\mathrm{p}$ & \imrppCHIPLIMITPerihelion & \seobnrCHIPLIMITPerihelion & \CHIPLIMITSYSPerihelion \\
\quad{}Primary spin magnitude $a_1$ & \imrppSPINONELIMITPerihelion & \seobnrSPINONELIMITPerihelion & \SPINONELIMITSYSPerihelion \\
\quad{}Secondary spin magnitude $a_2$ & \imrppSPINTWOLIMITPerihelion & \seobnrSPINTWOLIMITPerihelion & \SPINTWOLIMITSYSPerihelion \\
Lower bound  & & & \\
\quad{}Mass ratio $q$ & \imrppMASSRATIOLIMITPerihelion & \seobnrMASSRATIOLIMITPerihelion & \MASSRATIOLIMITSYSPerihelion \\
\rule{0pt}{3ex}%
Optimal SNR $\rho_{\langle h| h \rangle}$ & \imrppPEOPTIMALSNRPerihelion & \seobnrPEOPTIMALSNRPerihelion & \PEOPTIMALSNRPerihelion \\
Matched-filter SNR $\rho_{\langle h| s \rangle}$ & \imrppPEMATCHSNRPerihelion & \seobnrPEMATCHSNRPerihelion & \PEMATCHSNRPerihelion \\
\end{tabular}
\end{ruledtabular}
\label{tab:event-parameters}
\end{table*}

\begin{figure}
 \centering
 \includegraphics[width=\columnwidth]{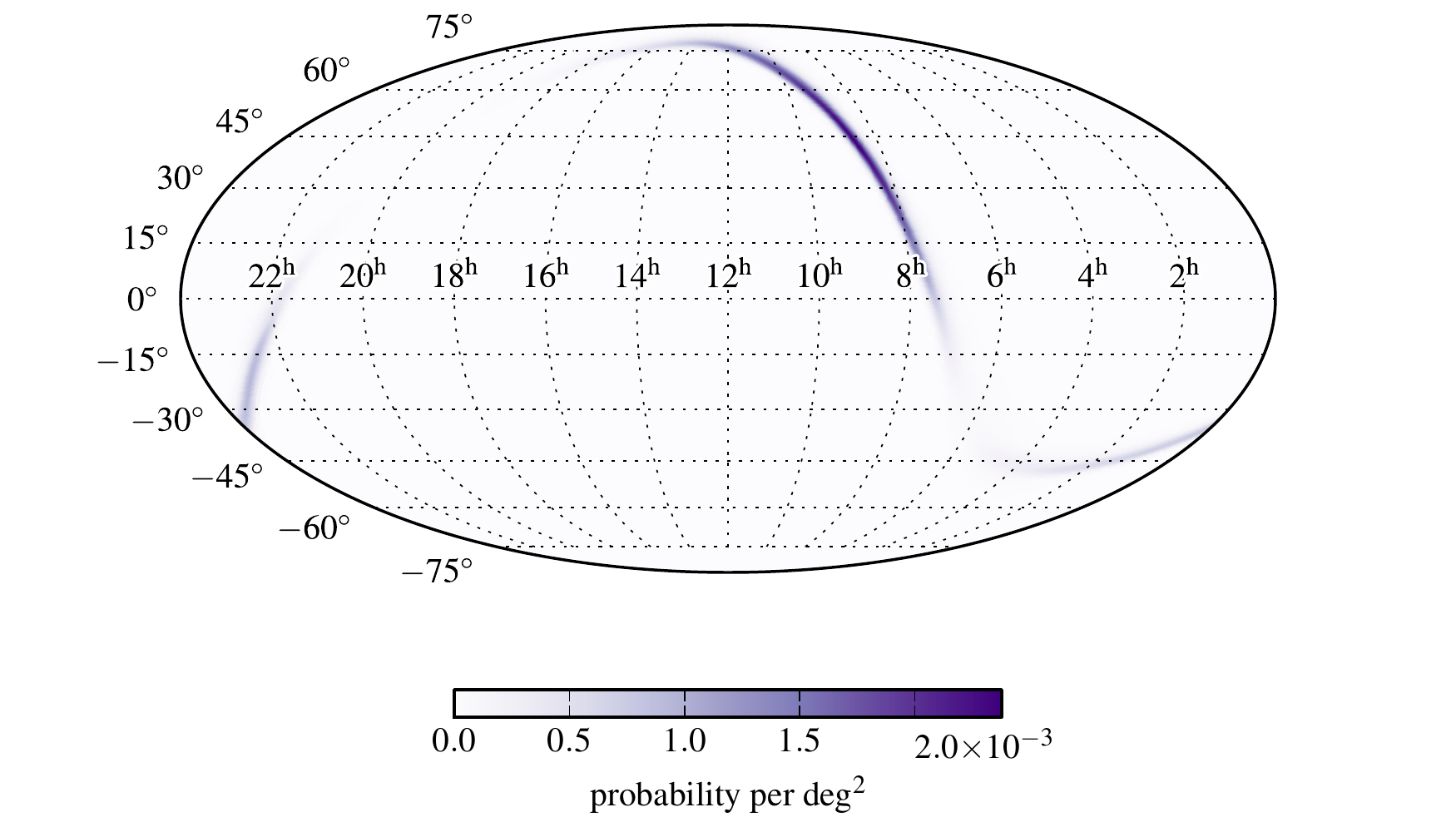}
 \caption{A Mollweide projection of the posterior probability density for the location of the source in equatorial coordinates (right ascension is measured in hours and declination is measured in degrees). The location broadly follows an annulus corresponding to a time delay of $\sim\imrppTIMEDELAYHLPerihelion~\mathrm{ms}$ between the Hanford and Livingston observatories. We estimate that the area of the $90\%$ credible region is $\sim\PESKYNINTYPerihelion$.
}
 \label{fig:sky}
\end{figure}

Figure~\ref{fig:distance} illustrates the distance, and the angle between the total angular momentum and the line of sight $\theta_{JN}$. 
The latter is approximately constant throughout the inspiral and serves as a proxy for the binary inclination~\cite{Apostolatos:1994mx,Farr:2014qka}. 
The full-precessing model shows a greater preference (after accounting for the prior) for face-on or face-off orientations with $\theta_{JN} \simeq 0^\circ$ or $180^\circ$. 
This leads to the tail of the $D_\mathrm{L}$ distribution extending to farther distances. 
There is a preference towards face-on or face-off inclinations over those which are edge on; the probability that $|\cos\theta_{JN}| > 1/\sqrt{2}$ is $\THETAJNPROBPerihelion$, compared to a prior probability of $0.29$. 
These inclinations produce louder signals and so are expected to be most commonly detected~\cite{Schutz:2011tw,Nissanke:2012dj}.  
Viewing the binary near face-on or face-off minimises the impact (if present) of precession~\cite{Vitale:2014mka,GW150914-PARAMESTIM}. 

\begin{figure}
 \centering
 \includegraphics[width=0.98\columnwidth]{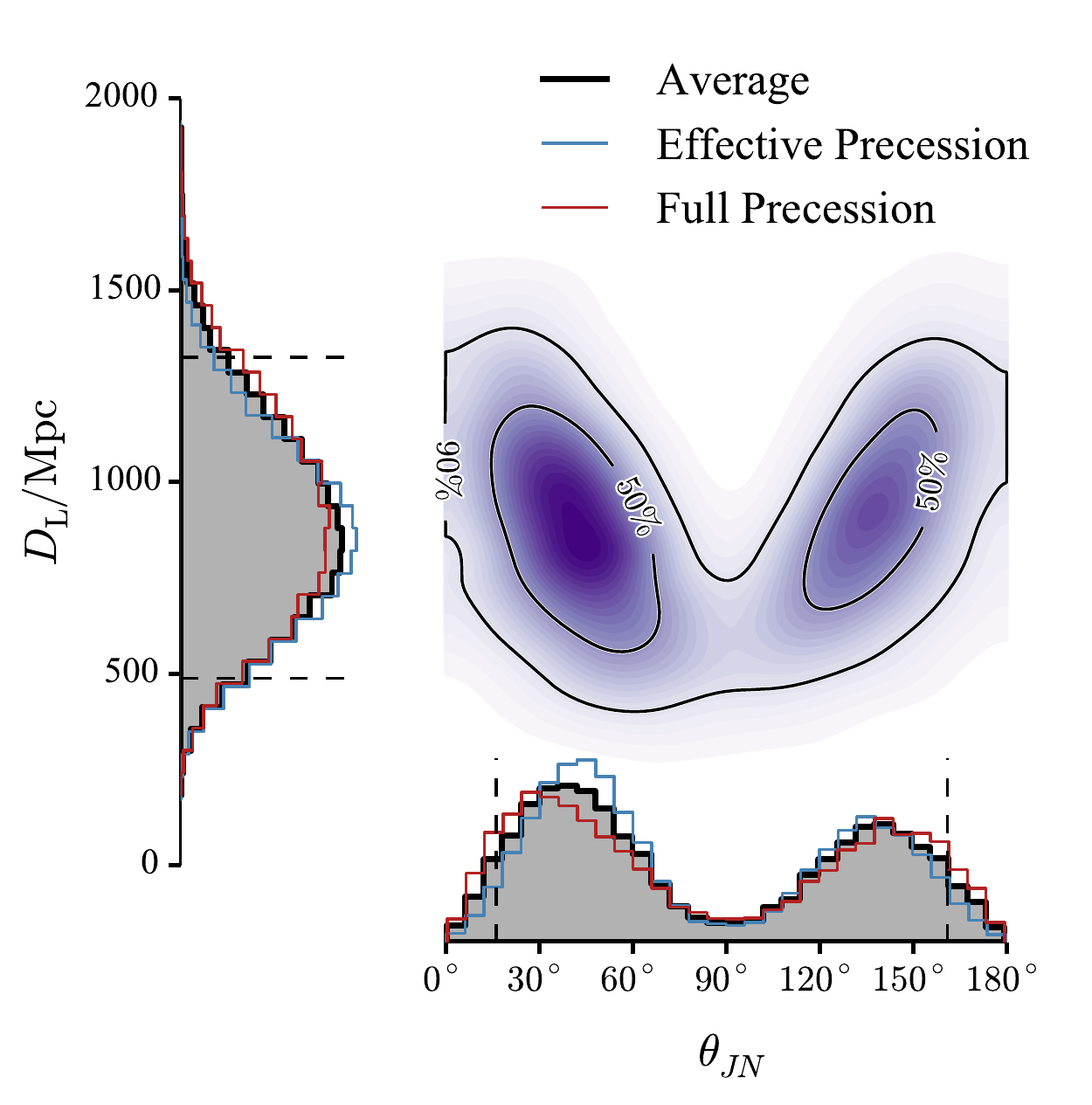}
 \caption{Posterior probability density for the source luminosity distance $D_\mathrm{L}$
and the binary inclination $\theta_{JN}$. 
The 
one-dimensional distributions include the posteriors 
for the two
waveform models, and their average (black). 
The dashed lines mark the $90\%$ credible interval for the average posterior.
The two-dimensional plot shows the $50\%$ and 
$90\%$ credible regions plotted over the posterior density function.
}
 \label{fig:distance}
\end{figure}

For \EventName{}, we obtain weak constraints on the spins. 
The amount of information we learn from the signal may be quantified by the Kullback--Leibler divergence, or relative entropy, from the prior to the posterior~\cite{Kullback:1951,Mackay:2003}. 
For $\chi_\mathrm{eff}$ we gain $\CHIEFFKLPerihelion~\mathrm{nat}$ of information, and for $\chi_\mathrm{p}$ we only gain $\CHIPKLPerihelion~\mathrm{nat}$. 
As comparison, the Kullback--Leibler divergence between two equal-width normal distributions with means one standard deviation apart is $0.5~\mathrm{nat} = 0.72~\mathrm{bit}$. 
We cannot gain much insight from these spin measurements, but this may become possible by considering the population of binary black holes~\cite{Mandel:2009nx}. 
Figure~\ref{fig:chi-all} shows the inferred $\chi_\mathrm{eff}$ distributions for \EventName, GW150914, LVT151012 and GW151226~\cite{O1:BBH}. 
Only GW151226 has a $\chi_\mathrm{eff}$ (and hence at least one component spin) inconsistent with zero. 
The others are consistent with positive or negative effective inspiral spin parameters; 
the probabilities that $\chi_\mathrm{eff} > 0$ are $\CHIEFFPROBPerihelion$, $\CHIEFFPROB$ and $\CHIEFFPROBSecondMonday$ for \EventName{}, GW150914 and LVT151012, respectively. 
Future analysis may reveal if there is evidence for spins being isotropically distributed, preferentially aligned with the orbital angular momentum, or drawn from a mixture of populations~\cite{Vitale:2015tea,Stevenson:2017dlk,Talbot:2017yur,Farr:2017}.

\begin{figure}
 \centering
 \includegraphics[width=0.98\columnwidth]{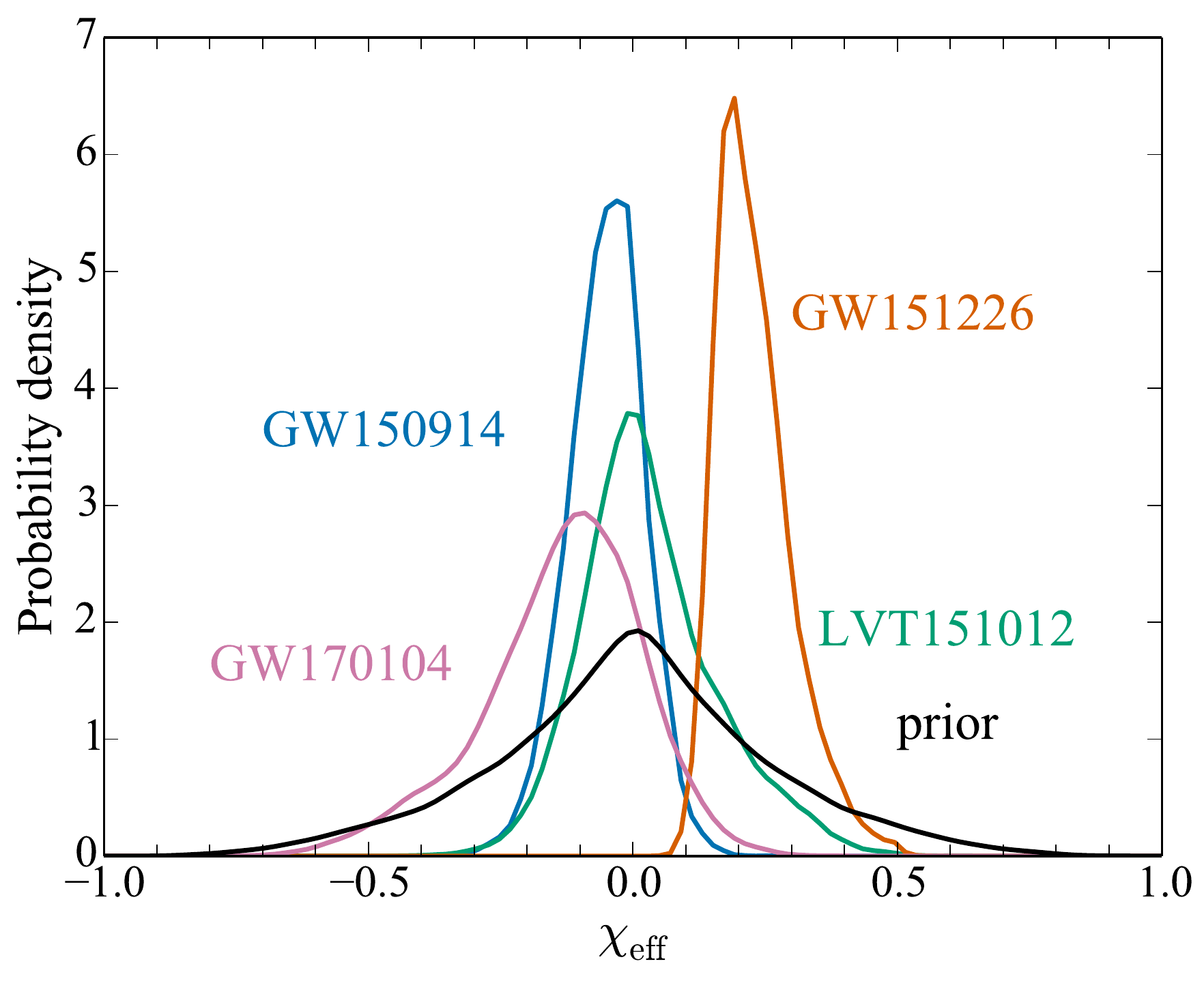}
 \caption{Posterior probability densities for the effective inspiral spin $\chi_\mathrm{eff}$ for \EventName, 
GW150914, LVT151012 and GW151226~\cite{O1:BBH}, 
together with the prior probability distribution for \EventName.
The distribution for \EventName{} uses both precessing waveform models, but, for ease of comparison, the others 
use only the effective-precession model. 
The prior distributions vary between events, as a consequence of different mass ranges, 
but the difference is negligible on the scale plotted.
}
 \label{fig:chi-all}
\end{figure}

While we learn little about the component spin magnitudes, we can constrain the final black hole spin. 
The final black hole's dimensionless spin $a_\mathrm{f}$ is set by the binary's total angular momentum 
(the orbital angular momentum and the components' spins) minus that radiated away.  
Figure~\ref{fig:final} illustrates the probability distributions for the final mass and spin. 
To obtain these, we average results from different numerical-relativity calibrated fits  
for the final mass~\cite{Jimenez-Forteza:2016oae,Healy:2016lce} and for the final spin~\cite{Hofmann:2016yih,Jimenez-Forteza:2016oae,Healy:2016lce}. 
The fitting formulae take the component masses and spins as inputs.
We evolve the spins forward from the $20~\mathrm{Hz}$ reference frequency to a fiducial near merger frequency, 
and augment the aligned-spin final spin fits~\cite{Jimenez-Forteza:2016oae,Healy:2016lce} with the contribution from the in-plane spins before averaging~\cite{T1600168}. 
From comparisons with precessing numerical-relativity results, we estimate that systematic errors are negligible compared to the statistical uncertainties. 
We follow the same approach for fits for the peak luminosity~\cite{Healy:2016lce,Keitel:2016krm}; 
here the systematic errors are larger (up to $\sim10\%$) and we include them in the uncertainty estimate~\cite{GW150914-PARAMESTIM}. 
We find that $a_\mathrm{f} = \SPINFINALinplaneCOMPACTPerihelion$.
This is comparable to that for previous events~\cite{GW150914-PARAMESTIM,GW150914-PRECESSING,O1:BBH}, as expected for near equal-mass binaries~\cite{Gonzalez:2006md,Berti:2007fi}, but extends to lower values because of the greater preference for spins with components antialigned with the orbital angular momentum.

\begin{figure}
 \centering
 \includegraphics[width=0.98\columnwidth]{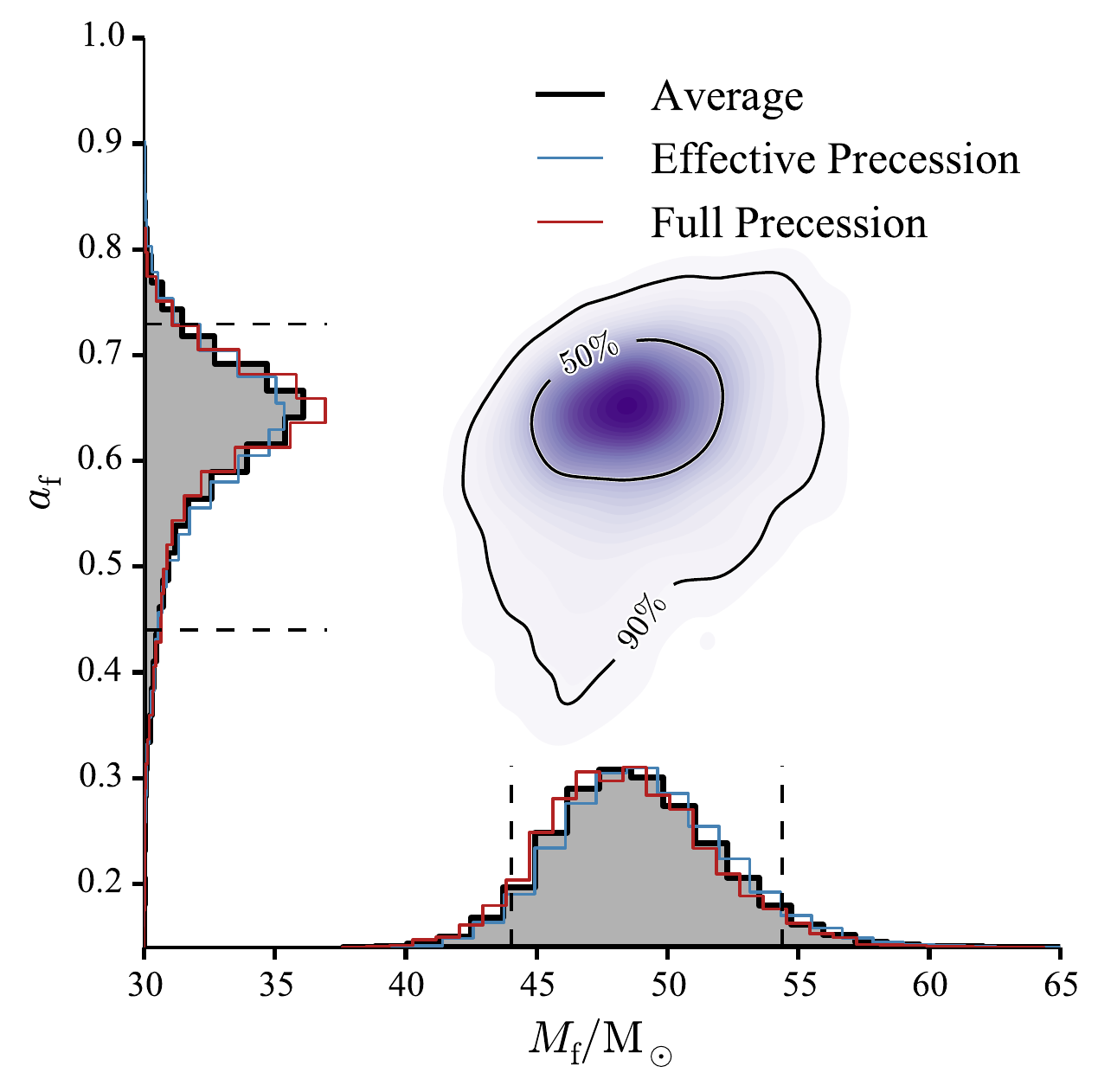}
 \caption{Posterior probability density for the final black hole mass $M_\mathrm{f}$ 
and spin magnitude $a_\mathrm{f}$. 
The 
one-dimensional distributions include the posteriors 
for the two
waveform models, and their average (black). 
The dashed lines mark the $90\%$ credible interval for the average posterior.
The two-dimensional plot shows the $50\%$ and 
$90\%$ credible regions plotted over the posterior density function.
}
 \label{fig:final}
\end{figure}

The final calibration uncertainty is sufficiently small to not significantly affect results. 
To check the impact of calibration uncertainty, we repeated the analysis using the effective-precession waveform without marginalising over the calibration. 
For most parameters the change is negligible. The most significant effect of calibration uncertainty is on sky localization. 
Excluding calibration uncertainty reduces the $90\%$ credible area by $\sim\SKYCALIBNINTYimrppPerihelion$.

  \section{Population inference}
\label{s:pop-inf}
Gravitational-wave observations are beginning to reveal a population of merging binary black holes. 
With four probable mergers we can only roughly constrain the population. 
Here we fit a hierarchical single-parameter population model to the three probable mergers from
first observing run~\cite{O1:BBH} and \EventName{}. 
We assume that the two-dimensional mass distribution of mergers is the combination of 
a power law in $m_1$ and a flat $m_2$ distribution,
\begin{equation}
  \label{eq:mass-dist-assumption}
  p\left( m_1 , m_2 \right) \propto m_1^{-\alpha}
  \frac{1}{m_1 - \MMin},
\end{equation}
with $\MMin = 5 \, \Msun$~\cite{Ozel:2010su,Farr:2010tu,Kreidberg:2012ud}, and subject to
the constraint that $M \leq \MMax$, with $\MMax = 100 \, \Msun$, 
matching the analysis from the first observing run~\cite{GW150914-RATES,O1:BBH}. 
Our sensitivity to these choices for lower and upper cut-off masses is much smaller
than the statistical uncertainty in our final estimate of $\alpha$.
The marginal distribution for $m_1$ is
\begin{equation}
  \label{eq:m1-distribution}
  p\left( m_1 \right) \propto m_1^{-\alpha} \frac{\min\left( m_1,
      \MMax - m_1 \right) - \MMin}{m_1 - \MMin},
\end{equation}
and the parameter $\alpha$ is the power-law slope of the marginal
distribution for $m_1$ at masses $m_1 \leq \MMax / 2$.  
The initial mass function of stars follows a similar power-law distribution~\cite{Salpeter:1955it,Kroupa:2000iv}, 
and the mass distribution of companions to
massive stars appears to be approximately uniform in the mass ratio $q$~\cite{Kobulnicky:2007,Sana:2012px,Kobulnicky:2014ita}. 
While the initial--final mass relation in binary black hole systems is
complicated and nonlinear~\cite{Fryer:1999ht,Fryer:2011cx,Dominik:2012kk,Spera:2015vkd}, 
this simple form provides a sensible starting point for estimating the mass distribution.

Accounting for selection
effects and the uncertainty in our estimates of the masses of our four
events, and imposing a flat prior on the parameter $\alpha$~\cite{O1:BBH}, we find
$\alpha = \pmexpintr$. 
Our posterior on $\alpha$ appears in
Fig.~\ref{fig:alpha-posterior}. The inferred posterior on the
marginal distribution for $m_1$ appears in Fig.~\ref{fig:m1-dist}; the
turnover for $m_1 > 50 \, \Msun$ is a consequence of our choice of
$\MMax = 100 \, \Msun$ in Eq.~\eqref{eq:m1-distribution}.
\begin{figure}
  \centering
  \includegraphics[width=0.995\columnwidth]{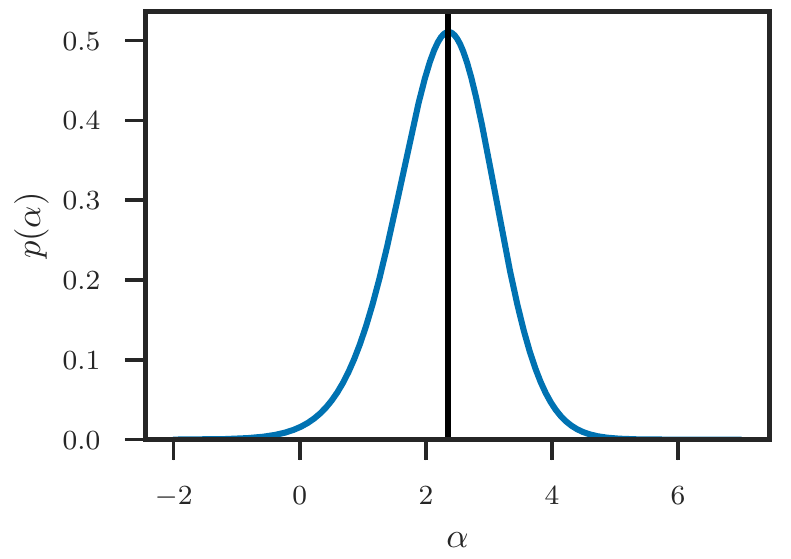}
  \caption{The posterior distribution for the power-law slope of the
    massive component of the binary black hole mass distribution, $\alpha$,
    described in the main text, using the three probable events from
    the first observing run~\cite{O1:BBH} and \EventName{}.  We find the median and $90\%$
    credible interval are $\alpha = \pmexpintr$.  The black line
    indicates the Salpeter law~\cite{Salpeter:1955it} slope used in the
    power-law population for estimating binary black hole coalescence rates.}
  \label{fig:alpha-posterior}
\end{figure}

\begin{figure}
  \centering
  \includegraphics[width=0.995\columnwidth]{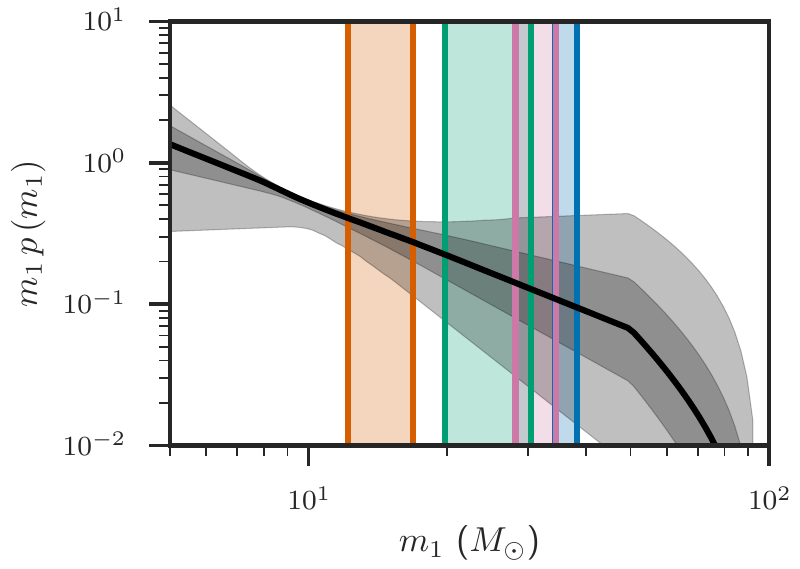}
  \caption{The posterior probability distribution 
    for the primary component mass $m_1$ of binary black holes
    inferred from the hierarchical analysis.
    The black line
    gives the posterior median as a function of mass, and the dark and
    light grey bands give the $50\%$ and $90\%$
    credible intervals.  The colored vertical bands give the $50\%$
    credible interval from the posterior on $m_1$ from the analyses of
    (left to right) GW151226, LVT151012, \EventName{}, and GW150914.  The marginal mass
    distribution is a power law for $m_1 \leq 50 \, \Msun$, and turns
    over for $m_1 \geq 50 \, \Msun$ due to the constraint on the
    two-dimensional population distribution that
    $m_1+m_2 \leq 100 \, \Msun$.}
  \label{fig:m1-dist}
\end{figure}

  \section{Tests of general relativity}
\label{s:tgr}
The tests of GR use the same algorithm base described in Sec.~\ref{s:pe-appendix}~\cite{Veitch:2014wba} 
for estimation of source parameters, with appropriate modifications to the analytical 
waveform models~\cite{GW150914-TESTOFGR,O1:BBH}.  In the Fourier domain, gravitational waves from a 
coalescing binary can be described by 
\begin{equation}
\tilde{h}_\mathrm{GR}(f) = \tilde{A}(f;\vec\vartheta_\mathrm{GR}) e^{i \Psi(f;\vec\vartheta_\mathrm{GR})},
\end{equation}
where $\vec\vartheta_\mathrm{GR}$ are the parameters of the source (e.g., masses and spins) in GR. 
The tests of GR we perform, except for the inspiral--merger--ringdown consistency test, introduce a dephasing term with an unknown prefactor that captures the magnitude of the deviation from GR. While we modify the phase of the waveform from its GR value, the amplitude is kept unchanged; this is because our 
analysis is more sensitive to the phase evolution than the amplitude.  We use a non-GR template of the form
\begin{eqnarray}
\tilde{h}(f) & = & \tilde{A}(f;\vec\vartheta_\mathrm{GR}) e^{i\left[\Psi(f;\vec\vartheta_\mathrm{GR})
 + \delta\Psi(f;\vec\vartheta_\mathrm{GR},X_\mathrm{modGR})\right]},
\end{eqnarray}
where $X_\mathrm{modGR}$ is a theory-dependent parameter, which is zero in the usual GR templates.  To
simulate the non-GR waveform, we used the effective-precession model as a base;
all the GR and non-GR parameters are assumed unknown and estimated from the data.

With multiple detections it is possible to combine constraints on $X_\mathrm{modGR}$ to obtain tighter bounds. 
For a generic parameter $\vartheta$, we compute a combined posterior distribution by combining the individual likelihoods~\cite{Mandel:2009pc}. For each event $e_i$ we estimate the marginal likelihood density $p(e_i | \vartheta)$ using a Gaussian kernel density estimator. This gives a simple representation of the likelihood that can be easily manipulated. 
The combined posterior distribution is computed by multiplying the marginal likelihoods and the chosen prior distribution, 
\begin{equation}
p(\vartheta|e_1,\ldots,e_N) \propto p(\vartheta)\prod_{i=1}^N p(e_i|\vartheta).
\end{equation}
This is used to compute bounds on $\vartheta$ given $N$ detections. 
We use the three confident detections (GW150914, GW151226 and \EventName{}) to set combined bounds on potential deviation from GR, except in the case of the inspiral--merger--ringdown consistency test 
where only GW150914 and \EventName{} are used as 
GW151226 has insufficient SNR from the merger--ringdown to make useful inferences.

\subsection{Modified dispersion}

We have assumed a generic dispersion relation of the form
$E^2=p^2 c^2+A p^{\alpha} c^\alpha$, $\alpha\geq0$.  
To leading order in $A E^{\alpha-2}$, the group velocity of gravitational waves is thus modified as $v_g/c = 1 + (\alpha -1) A E^{\alpha-2} /2$. 
The modified dispersion relation results in an extra term to be added to the gravitational-wave 
phase~\cite{Mirshekari:2011yq}:
\begin{equation}\label{Eq.LIVdPsi}
\delta\Psi = 
\begin{cases}
\displaystyle \frac{\pi}{\alpha-1} \frac{A D_{\alpha}}{(h c)^{2-\alpha}} \left[\frac{(1+z)f}{c}\right]^{\alpha-1} & \alpha \neq 1 \\[15pt]
\displaystyle \frac{\pi A D_\alpha}{h c} \ln\left(\frac{\pi G \mathcal{M}^\mathrm{det} f}{c^3}\right) & \alpha = 1
\end{cases}.
\end{equation}
Here $\mathcal{M}^\mathrm{det}$ is the redshifted (detector-frame) chirp mass, $h$ is the Planck constant, and $D_{\alpha}$ is a distance 
measure,
\begin{equation}
D_\alpha = \frac{1+z}{H_0} \int_{0}^{z} \frac{(1+z')^{\alpha - 2}}{\sqrt{\Omega_\mathrm{m}(1+z')^3 + \Omega_\Lambda}}\,\mathrm{d}z',
\end{equation}
where $H_0$ is the Hubble constant, $\Omega_\mathrm{m}$ and $\Omega_\Lambda$ are the matter and dark energy density parameters~\cite{Ade:2015xua}, respectively.

\begin{table*}[htb]
\caption{$90\%$ credible level upper bounds on the Lorentz violation magnitude $|\lora{}/\mathrm{eV}^{2-\alpha}|$ using GW150914, GW151226, \EventName, and their joint posterior.}\label{Table:Liv_Combined}
\begin{ruledtabular}
\begin{tabular}{ l c c c c c  c c c c }
\centering
         & \multicolumn{4}{c}{$A>0$} & & \multicolumn{4}{c}{$A<0$} \\
\cline{2-5} \cline{7-10}
\multicolumn{1}{c}{$\alpha$} & GW150914 &GW151226 & \EventName{} & Joint & &  GW150914 &GW151226 & \EventName{} & Joint \\\hline
$0.0$ &$1.3\times 10^{-44}$ & $1.7\times 10^{-43}$ & $9.8\times 10^{-45}$ &$7.3\times 10^{-45}$ &&$2.3\times 10^{-44}$ & $7.1\times 10^{-44}$ & $3.6\times 10^{-44}$ & $1.8\times 10^{-44}$\\ 
$0.5$ & $4.8\times 10^{-38}$ &$1.6\times 10^{-37}$ &    $1.8\times 10^{-38}$ & $1.7\times 10^{-38}$ &&$4.1\times 10^{-38}$ & $9.4\times 10^{-38}$ &$7.8\times 10^{-38}$ & $2.9\times 10^{-38}$\\
$1.0$ & $8.5\times 10^{-32}$ & $1.8\times 10^{-31}$ &   $3.6\times 10^{-32}$ & $2.8\times 10^{-32}$ &&$1.0\times 10^{-31}$&$1.3\times 10^{-31}$ & $1.0\times 10^{-31}$ & $5.0\times 10^{-32}$\\
$1.5$ & $1.9\times 10^{-25}$ &$3.2\times 10^{-25}$ &    $9.4\times 10^{-26}$ & $7.5\times 10^{-26}$ &&$2.7\times 10^{-25}$ &$2.2\times 10^{-25}$ &$2.3\times 10^{-25}$ & $1.1\times 10^{-25}$\\

$2.5$ & $3.9\times 10^{-13}$ &$1.4\times 10^{-13}$ &    $2.8\times 10^{-13}$ &$1.2\times 10^{-13}$ &&$2.8\times 10^{-13}$ &$2.0\times 10^{-13}$ & $1.3\times 10^{-13}$ & $8.9\times 10^{-14}$\\
$3.0 $&$2.2\times 10^{-07}$ & $7.4\times 10^{-08}$ &    $1.7\times 10^{-07}$ &$6.2\times 10^{-08}$ &&$1.7\times 10^{-07}$ &$1.5\times 10^{-07}$ & $8.9\times 10^{-08}$ & $4.3\times 10^{-08}$\\
$3.5$ &  $1.7\times 10^{-01}$ &$5.4\times 10^{-02}$ &     $1.4\times 10^{-01}$ & $4.2\times 10^{-02}$ &&$1.2\times 10^{-01}$ &$1.1\times 10^{-01}$ & $7.1\times 10^{-02}$ &$2.6\times 10^{-02}$\\
$4.0$ & $1.3\times 10^{+05}$ &$5.9\times 10^{+04}$ &      $1.0\times 10^{+05}$ & $2.8\times 10^{+04}$&& $9.7\times 10^{+04}$ & $1.3\times 10^{+05}$ & $7.7\times 10^{+04}$ &$2.0\times 10^{+04}$\\
\end{tabular}
\end{ruledtabular}
\end{table*}

Table~\ref{Table:Liv_Combined} lists the $90\%$ credible upper bounds on the magnitude of $\lora$, 
where the individual and combined bounds for the three confident
detections are shown; we see that depending on the value of $\alpha$ and the sign of $\lora$, the
combined bounds are better than those obtained from \EventName{} alone by a factor of $\sim1$--$4.5$.
For all values of $\alpha$, these bounds are consistent with the uncertainties one might expect for heavy binary black holes using Fisher-matrix estimates on simulated GW150914-like signals~\cite{Yunes:2016jcc}.

For small values of $\alpha$, it is useful to recast the results in terms of lower bounds on a length scale $\lambda_\lora = h c A^{1/(\alpha -2)}$, which can be thought of as the range (or the screening length) of an effective potential, which is infinite in GR. In Table~\ref{Tab.LIBLambda} we report the numerical values of these bounds for $\alpha<2$. 
For $\alpha=3,4$, we instead express the bounds as lower limits on the energy scale at which quantum gravity effects might become important, $E_\mathrm{QG} = A^{-1/(\alpha-2)}$~\cite{Moore:2001bv,Borriello:2013ala,Kiyota:2015dla,Wei:2016exb,Wang:2016lne}.
This facilitates the comparison with existing constraints from other sectors, which we show in Table~\ref{Table:LivEQG}. 
\begin{table}[htb]
\caption{$90\%$ credible level lower bounds on the length scale $\lambda_\lora$ for Lorentz invariance violation test using \EventName{} alone.}\label{Tab.LIBLambda}
\begin{ruledtabular}
\begin{tabular}{ c l l }
& \multicolumn{1}{c}{$\lora >0$} & \multicolumn{1}{c}{$\lora<0$} \\\hline
$\alpha=0.0$ & $1.3\times 10^{13}~\mathrm{km}$ & $6.6\times 10^{12}~\mathrm{km}$\\
$\alpha=0.5$ & $1.8\times 10^{16}~\mathrm{km}$ & $6.8\times 10^{15}~\mathrm{km}$\\
$\alpha=1.0$ & $3.5\times 10^{22}~\mathrm{km}$ & $1.2\times 10^{22}~\mathrm{km}$\\
$\alpha=1.5$ & $1.4\times 10^{41}~\mathrm{km}$ & $2.4\times 10^{40}~\mathrm{km}$\\
\end{tabular}
\end{ruledtabular}
\end{table}

\newcommand*\rot{\rotatebox{90}}
\begin{table}
\caption{$90\%$ confidence level lower bounds on the energy scale at which quantum gravity effects might become important $E_\mathrm{QG}$. Bounds are grouped into theories which produce subluminal and superluminal gravitational-wave propagation. The results from \EventName{} are considerably less constraining than those obtained with other methods, but they are the first direct constraints of Lorentz invariance violation in the superluminal gravity sector.}\label{Table:LivEQG}
\begin{ruledtabular}
\begin{tabular}{c l c c}
&& $\alpha =3$ & $\alpha =4$ \\\hline
\multirow{4}{*}{\rot{{Sub}}}  & \EventName{} & $1.1 \times 10^{7}~\mathrm{eV}$ & $3.6\times 10^{-3}~\mathrm{eV}$ \\
& Gamma rays~\cite{Wei:2016exb} & $5 \times 10^{24}~\mathrm{eV}$ & $1.4 \times 10^{16}~\mathrm{eV}$\\
& Neutrino~\cite{Wang:2016lne} & $1.2 \times 10^{26}~\mathrm{eV}$ & $7.3 \times 10^{20}~\mathrm{eV}$ \\
& Cherenkov~\cite{Moore:2001bv,Kiyota:2015dla} & $4.6\times 10^{35}~\mathrm{eV}$ & $5.2\times10^{27}~\mathrm{eV}$ \\
\hline 
\multirow{2}{*}{\rot{{Super}}} & \EventName{} & $6.0 \times 10^6~\mathrm{eV}$ & $3.2 \times 10^{-3}~\mathrm{eV}$ \\
 & Neutrino~\cite{Borriello:2013ala} & $1.2 \times 10^{33}~\mathrm{eV}$ & $1.2 \times 10^{24}~\mathrm{eV}$ \\
\end{tabular}
\end{ruledtabular}
\end{table}

In the subluminal propagation regime, bounds exist from electromagnetic (spectral time lag in
gamma-ray bursts~\cite{Wei:2016exb}), neutrino (time delay between neutrino and photons from blazar PKS
B1424-418~\cite{Wang:2016lne}), and gravitational (absence of gravitational Cherenkov
radiation~\cite{Moore:2001bv,Kiyota:2015dla}) sectors. 
In the superluminal propagation regime, the only existing limits are from the neutrino sector (absence of Bremsstrahlung
from electron--positron pairs~\cite{Borriello:2013ala}). The \EventName{} constraints are weaker than existing bounds, but are
the first constraints on Lorentz violation in the gravitational superluminal-propagation sector.

The posterior distributions for \lora{} have long tails, which makes it difficult to accurately calculate $90\%$ limits with a finite number of samples. 
To quantify this uncertainty on the bounds, 
for each value of $\alpha$ and sign of \lora{} we use Bayesian bootstrapping~\cite{Rubin:1981} to generate $1000$ instances of the relevant posterior distribution. 
We find that the $90\%$ credible upper bounds are estimated within an interval whose $90\%$ credible interval width is $\lesssim20\%$ of the values reported in Table~\ref{Table:Liv_Combined}.

For the (GR) source parameters, to check for the potential impact of errors from waveform modelling, we analysed the data using both the effective-precession model and the full-precession model. 
However, the full-precession model was not adapted in time for tests of GR to be completed for this publication. 
In the first observing run, we performed tests with two different waveform families~\cite{GW150914-TESTOFGR,O1:BBH}: the effective-precession model~\cite{Husa:2015iqa,Khan:2015jqa,Hannam:2013oca}, and a nonprecessing waveform model~\cite{Taracchini:2013rva,Purrer:2014fza}. 
We follow the same approach here, and use the same nonprecessing waveform model used for the matched filter search~\cite{Bohe:2016gbl}. The use of a nonprecessing waveform should give conservative bounds on the potential error from waveform modelling, as some of the differences may come from the failure to include precession effects~\cite{GW150914-PRECESSING}. 
We find that the numbers so obtained are consistent with the results of the effective-precession model at the tens of percent level. 

\subsection{Parametrized test}

\begin{figure*}
 \centering
 \includegraphics[width=\textwidth]{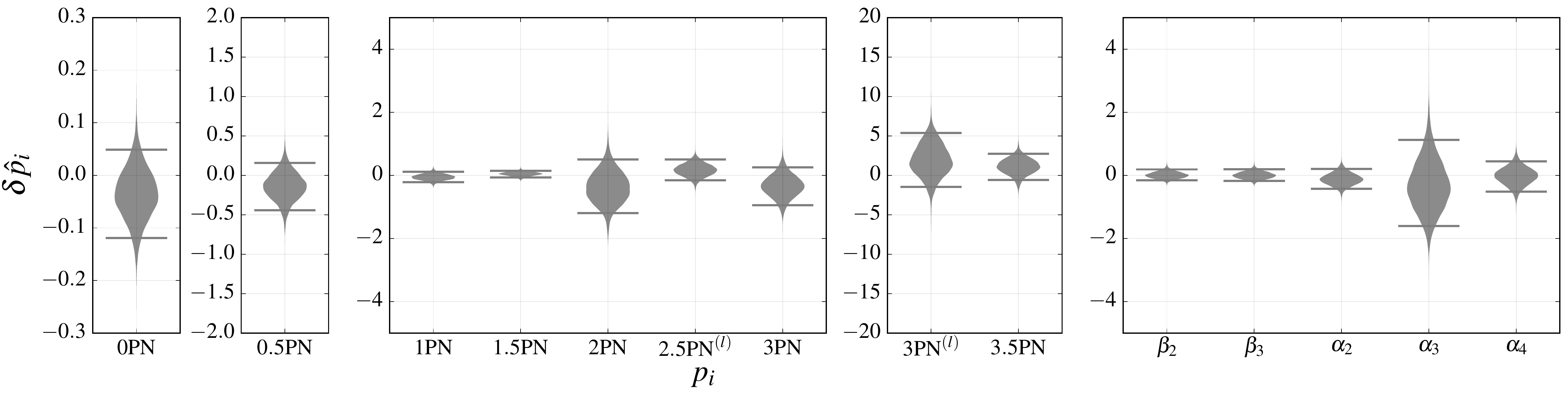}
 \caption{Violin plots for the parametrized test, combining posteriors for \EventName{} with the two confident detections made in the first observing run, GW150914 and GW151226~\cite{O1:BBH}. This figure has been corrected according to \cite{GW170104-Erratum}, and its 90\% bars have been re-positioned to correct a previous plotting error.}
 \label{fig:tgr_parametrized_combined}
\end{figure*}

The phase evolution of gravitational waves from compact binaries is well understood within GR. The
inspiral portion, corresponding to large orbital separation, can be described analytically using the
post-Newtonian expansion~\cite{Blanchet:2013haa}.  Modelling the merger dynamics requires the use of
numerical-relativity simulations~\cite{Pretorius:2005gq,Campanelli:2005dd,Baker:2006ha}, whereas the
post-merger signal is described in black hole perturbation theory as a superposition of damped
sinusoids~\cite{Vishveshwara:1970zz,Chandrasekhar:1975zza,Teukolsky:1973ha,Kokkotas:1999bd}.
Accurate analytical waveforms are obtained by tuning the
effective-one-body~\cite{Buonanno:1998gg,Buonanno:2000ef,Bohe:2016gbl} or phenomenological models 
\cite{Ajith:2007qp,Khan:2015jqa} to numerical-relativity 
simulations~\cite{Mroue:2013xna,Husa:2015iqa,Chu:2015kft}.

Given a phase parameter in the phenomenological model whose value in
GR is $p_i,$ we modify the waveform by introducing new
dimensionless parameters $\delta \hat p_i$ such that $p_i \rightarrow
p_i (1+\delta \hat p_i)$~\cite{GW150914-TESTOFGR,O1:BBH}.  In the parametrized null test, 
we freely vary one $\delta\hat p_i$ at a time (in addition to the other source parameters) 
to look for deviations from GR.

The bounds on $\hat p_i$ obtained from \EventName{} are weaker than those from the two confident detections of the first observing run~\cite{O1:BBH}. 
GW151226 had an SNR comparable to \EventName{}, but it is from a significantly lower mass system~\cite{GW151226-DETECTION,O1:BBH}, and hence places better constraints on the inspiral parameters.
GW150914 had an SNR twice that of \EventName{} (while being of comparable mass), and thus places the best constraints on the late-inspiral and merger--ringdown parameters.  
Therefore, instead of reporting bounds from \EventName, we provide updated combined bounds, combining the results from the three events. In Fig.~\ref{fig:tgr_parametrized_combined} we show a violin plot for each of the test parameters. 
The parameters are plotted (from the left) following the order in which they appear in the post-Newtonian expansion or enter the phenomenological model (the $\beta$ and $\alpha$ parameters). For all the parameters, the GR solution ($\delta\hat p_i = 0$) is contained in the $90\%$ credible interval.

\subsection{Inspiral--merger--ringdown consistency test}

GR is well tested in weak gravitational fields, but fewer tests have been performed in the strong-field regime~\cite{Will:2014kxa,Psaltis:2008bb,Berti:2015itd}. 
It is possible that deviations from the expected behavior of GR only manifest in the most extreme conditions, where spacetime is highly dynamical. 
The inspiral--merger--ringdown consistency test checks whether the low-frequency, inspiral-dominated portion of the waveform is consistent with the high-frequency, merger--ringdown portion. 
The two frequency ranges are analysed separately, and the inferred parameters are compared. 
The test uses the estimated final black hole mass and spin (calculated from the component masses and spins using numerical-relativity fits as detailed in Sec.~\ref{s:pe-appendix})~\cite{GW150914-TESTOFGR,Ghosh:2016qgn}. 
If the waveform is compatible with the predictions of GR, we expect that the parameters inferred from the two pieces will be consistent with each other, although the difference will not, in general, be zero because of detector noise.
\begin{figure}[!hbt]
 \includegraphics[width=0.95\columnwidth]{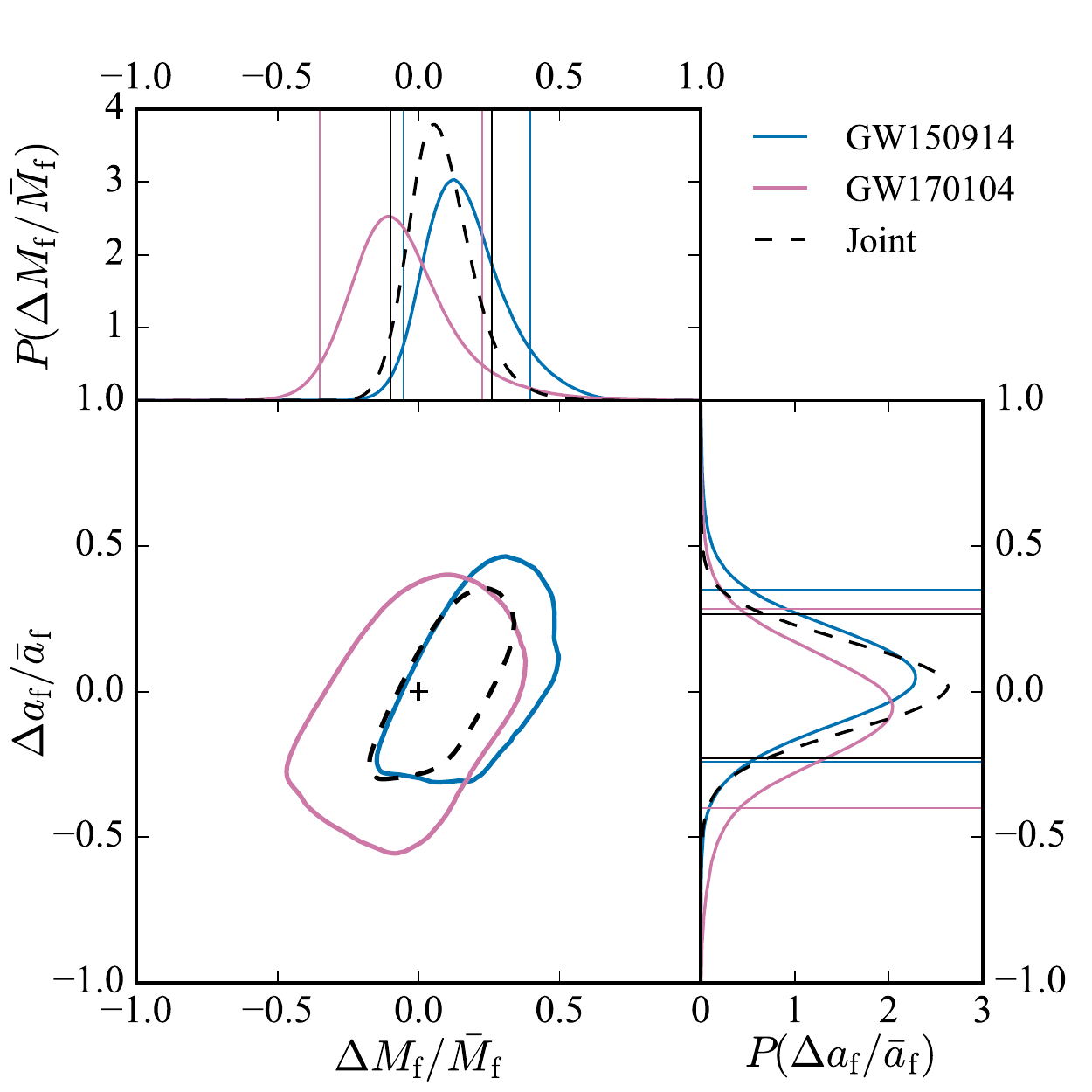}
 \caption{Posterior probability distributions for the fractional differences in the remnant black hole mass $\Delta M_\mathrm{f}/\bar{M}_\mathrm{f}$ and spin $\Delta a_\mathrm{f}/\bar{a}_\mathrm{f}$ calculated using the low-frequency (inspiral) and high-frequency (merger--ringdown) parts of the waveform. The GR solution is at $(0,0),$ shown in the two-dimensional plot as a black $\mathbf{+}$ marker. The contours show the $90\%$ credible region, the lines in the one-dimensional histograms mark the $90\%$ credible interval. We show the posteriors for \EventName{} and GW150914, as well as the combined posterior using both.}
 \label{Fig.TGR.IMR}
\end{figure}
In Fig.~\ref{Fig.TGR.IMR}, we show the posteriors on the fractional difference in the two estimates of the final mass and spin  for \EventName{} and GW150914, as well as the combined posterior.
The difference in the estimates are divided by the mean of the two estimates to produce the fractional parameters that describe potential departures from the GR predictions: $\Delta a_\mathrm{f}/\bar{a}_\mathrm{f}$ for the spin and $\Delta M_\mathrm{f}/\bar{M}_\mathrm{f}$ for the mass~\cite{Ghosh:2017gfp}. 
These definitions are slightly different from the ones used in our earlier papers~\cite{GW150914-TESTOFGR,Ghosh:2016qgn}, but serve the same qualitative role~\cite{Ghosh:2017gfp}. 
Each of the distributions is consistent with the GR value. The posterior for \EventName{} is broader, consistent with this event being quieter, and having a lower total mass, which makes it harder to measure the post-inspiral parameters.
The width of the $90\%$ credible intervals for the combined posteriors of $\Delta M_\mathrm{f}/\bar{M}_\mathrm{f}$ are smaller than those computed from GW170104 (GW150914) by a factor of $\sim 1.6 ~ (1.3)$, and the intervals for $\Delta a_\mathrm{f}/\bar{a}_\mathrm{f}$ are improved by a factor of $\sim 1.4 ~ (1.2)$.

}{%
}

\bibliography{GW170104}

\clearpage

\iftoggle{endauthorlist}{
  %
  %
  \let\author\myauthor
  \let\affiliation\myaffiliation
  \let\maketitle\mymaketitle
  \title{Authors}
  \pacs{}

  \newpage
  \maketitle
}

\end{document}